# Reducing spontaneous Raman scattering noise in high quantum bit rate QKD systems over optical fiber

Michal Mlejnek[1], Nikolay A. Kaliteevskiy[2], and Dan A. Nolan[1]

[1]Corning Research and Development Corporation, Corning, NY 14831, USA
[2]Corning Scientific Centre, Saint Petersburg, Russia

**Abstract:** We investigate (a) spontaneous Raman scattering of classical signals in long optical fiber links leading to significant Raman noise in quantum channels propagating in the same fiber and (b) impact of chromatic dispersion on the performance of high quantum bit rate QKD systems. One way of decreasing Raman noise is to decrease the classical signal power. This may be achieved by the use of coherent modulation formats and the deployment of Forward Error Correction technology. Additionally, low loss fiber with small Raman gain at telecom wavelengths enables the simultaneous transmission of multiple quantum channels and one or a few classical channels at 100 Gb/s rate over tens, possibly hundreds, of kilometers.

**Table of Contents**

# I. Introduction

Almost all new long-haul terrestrial and submarine communication systems deployed now use coherent transmission at 100 Gb/s or higher and no optical dispersion compensation within the optical link; i.e. chromatic dispersion is compensated digitally. Most current field studies of simultaneous quantum and classical signal transmission were performed in dark fiber, which is becoming scarce. The dark fiber also often does not provide the low loss and low optical nonlinearity medium required for efficient, high capacity modern coherent systems. Thus, it typically limits classical channel bit rates to very low values [1].

The understanding of the consequences and benefits of employing the new coherent systems and low loss, low nonlinearity fibers for classical channel transmission on the interaction with any additional quantum transmission channel over the same fiber is of interest. Hence we set out to study the quantum/classical channel wavelength-division multiplexing (WDM) in the context of quantum key distribution (QKD) [2],[3] in the presence of classical channels. To take full advantage of the low loss fiber, we consider the WDM channels placed in the transmission C-band from 1530 nm to 1565 nm.

QKD over a dedicated fiber has been shown to be a viable technology [4],[5],[6],[7],[8],[9],[10],[11],[12],[13]. In the case of fiber completely dedicated to QKD, secure key rates over 1 Mbps [5],[6],[7] and transmission distance over 300 km [12] and reaching 400 km [13] (not necessarily at the high secure key rates) were achieved. Most of the experiments were done on dark fiber, even though recently, low loss fibers were employed in the record setting QKD propagation distances [12],[13]. In either case, there is an effort to find conditions under which the fiber can be used for both QKD and classical signal transmission simultaneously. The main challenge is the possible contamination of a very weak QKD signal, which typically uses an average number of photons per pulse smaller than 1 (to preserve its "quantum nature") [14],[15], by a strong classical optical signal pulse containing on average approximately $10^4$ photons/pulse or more for a Gb/s link without amplification, and even more for higher classical bit rates [9]. While linear crosstalk can be battled by introducing spectral filtering, fiber optical nonlinearities cause inelastic scattering of photons that generates photons in the spectral region used by the QKD signal. Also note that, since the quantum signal is so weak, fiber with low loss will increase the information transfer over longer distance. (The classical channels can have ≤100 km spans, and with amplifiers reach 1000's of km.)

These general considerations lead one to study the effects of spontaneous Raman scattering (SpRS), four-way mixing (FWM) nonlinearity, and linear channel crosstalk (LCXT) on the quantum channel performance. All of these phenomena have been studied previously, e.g. [1],[8],[16]. The studies [1],[8],[16] used older, incoherent modulation formats to transmit the classical signal, which did not take an advantage of forward error correction (FEC). Here we update the study by considering modern coherent modulation formats (CMF) with FEC to study theoretically the impact of SpRS on the quantum channel in modern classical/quantum WDM transmission systems.

We shall not discuss in any detail the simple solutions that avoid the SpRS or LCXT: The use of time-division-multiplexing (TDM) or use of multi-core fibers (MCF) with dedicated cores either to QKD or classical channels.

The paper is organized as follows: In Section II the requirements/parameters of a basic transceiver setup are described. In Section 0 we describe the basic physical mechanism behind the various impairments of a QKD channel caused by the presence of an active classical channel within one (single-core) fiber – SpRS, FWM, and LCXT. We also discuss the "intrinsic" effect of fiber loss and chromatic dispersion on a QKD channel. Section IV discusses the advanced modulation formats with FEC impact on the QKD channel performance, characterized by quantum bit-error rate (QBER) and secure key rate. Detailed expressions for both characteristics are given for various QKD protocols. Several selected examples of QKD systems over a fiber with and without classical channels are evaluated in Section V, before the conclusions. We focus on a performance comparison of QKD systems over fibers similar to Corning® Vascade® EX2000 optical fiber, low dispersion Corning® LEAF® optical fiber, dispersion shifted or low dispersion fibers, and standard single mode fiber. Details about some classical detection aspects and chromatic dispersion of pulses are presented in the Appendices.

# II. Basic setup – parameters and requirements

The purpose of this paper is to theoretically determine the dependence of the performance of a classical/quantum WDM transmission on the fiber and classical channel modulation format parameters. As a practical example we chose the setup of Ref. [8], in which a Cerberis QKD system (from ID Quantique [17]) assumes four classical communication channels set up in addition to the quantum communication channel. Two classical channels are required for the bidirectional encrypted data transmission between the encryptors and an additional two classical authenticated channels are required for distillation, i.e. key sifting, error correction and privacy amplification, one from Alice to Bob and one from Bob to Alice.

### A. Quantum encoding parameters

The quantum transmission part is characterized by the frequency $f_{rep}$ with which Bob generates the pulses, or by the associated quantum bit period $T \equiv 1/f_{rep}$. The pulses are sent to Alice, who uses them for clock synchronization, attenuates them, modulates them, and sends them back as a quantum signal to Bob. In order to avoid compromising Rayleigh backscatter, the sequence length is chosen to match twice the length of the storage line of $l_A$ at Alice's [8] (the attenuation in the storage line is irrelevant, since it is applied before the quantum signal is attenuated to the desired average photon number in a pulse, $\mu$).

Avalanche photo diode (APD) detectors are used to detect the quantum signal. We use the quantum detector gate time (i.e., temporal filtering) $\Delta t_{gate}$, dead time $\tau_{dead}$, after-pulse probability to the total detection probability fraction $\rho_{AP}$, and quantum detection efficiency $\eta$ as characteristic parameters of APDs. If the quantum bit rate is not high and the chromatic dispersion may be neglected then we assume that $\Delta t_{gate}$ is on the order of or larger than the quantum pulse duration so as not to incur any additional loss due to clipping of the pulse in the time domain. In the case if the chromatic dispersion has a significant impact on QKD link performance the relation of the pulse duration and $\Delta t_{gate}$ should be optimized. Before being detected, the quantum signal experiences an internal loss $t_B$ due to optical components in Bob's detection system. The interference visibility $V$ is measured in order to help determining the QBER.

For the secret key distillation phase, the signals are sifted using a chosen QKD protocol (we shall consider here only BB84 [2] with optional filtering, and COW [17],[18] protocols). Following the key sifting is a fully implemented quantum error correction. We assume that its performance is similar to the CASCADE algorithm [18] as used in Ref. [8]. After privacy amplification using hashing functions based on Toeplitz matrices [20], Alice and Bob remain with shared secret keys.

In Ref. [8], for encrypted data transmission, a permanent advanced encryption standard AES-256 (key length = 256 bits, working with 128 bits long blocks [21]), encrypted 1 Gb/s data link between Alice and Bob was achieved using a pair of Ethernet encryptors. The encryptors were periodically updated with the secret keys with a typical key refresh rate once per minute. This requires a secret key rate of at least approximately 8.6 b/s (i.e., 4.3 b/s per encrypted classical channel). If the secret key rate drops below that limit, the key refresh rate was temporarily reduced to assure continuous operation. Standard coherent transponders have 25 Gbaud (with no FEC overhead, commercially 32 Gbaud with FEC overhead) information symbol rate that corresponds to 50 and 100 Gb/s capacity for PM-BPSK or PM-QPSK modulation formats, respectively. To keep the same relation of secret key refresh period to the size of encrypted data (i.e. keep the security level constant), the secret key rate threshold should be increased 50(100) times up to approximately 430(860) b/s for PM-BPSK(PM-QPSK). To preserve the amount of encrypted data sent during refresh period, e.g. 60 Gb data per 1 minute for the Ethernet system, the following strategies may be used: (a) increase the secret key rate threshold for one quantum channel, e.g. to 860 b/s for PM-QPSK, (b) increase the number of quantum channels, or (c) a combination of these methods. If the required secret key rate is unachievable, we may increase the secret key refresh period along with a corresponding decrease of the encryption reliability. There may be various restraints on the minimal secret key refresh period; currently about 60s is feasible [22], which would keep the secret key rate threshold at the ~4.3 b/s level for AES-256 per encrypted classical channel. More detail about the quantum signal preparation and transmission can be found in Ref. [8] and references therein.

### B. Classical channel characteristics

In our analysis we shall consider four or more classical communication channels that are implemented using standard optical 100 Gb/s WDM transceivers with varying CMF, such as PM-QPSK, DPSK, or 16QAM [23]. The corresponding receiver sensitivity is denoted by $R_x$ and it characterizes the minimum optical signal power incident on the detector such that the signal can be corrected and recovered with less than a given small final bit error ratio (BER); a common choice is $10^{-12}$ [21],[23].

Standard single-mode fiber of different span lengths is used as a fiber link of length $L$. The average fiber attenuation $\alpha$ is characterized in dB/km. For practical reasons, all classical channels are chosen to have larger wavelength than quantum to take advantage of a lower Raman noise on the anti-Stokes side of the Raman spectrum (see Section III.A below).

Any practical WDM system suffers from insertion loss (IL) and LCXT. The total IL $t_{IL} + t_{IL,FBG}$ takes into account optical filtering in the WDM ($t_{IL}$) as well as any spectral filtering introduced by additional components, or misalignments in the system ($t_{IL,FBG}$). LCXT depends on the isolation, $t_i$, of neighboring WDM channels provided by the WDM filters (typically having the width on the order of the bit rate in a classical channel). For generic purposes we shall assume that the adjacent 100G classical channels are distinguished by a 4[th] order Butterworth filter on a 100 GHz grid (0.8 nm apart) with isolation $t_{i,a}$, and width wide enough not to cause additional loss due to clipping of the spectrum of the quantum signal pulses. To lower the LCXT effect on the quantum channel while using the same WDM filter, the quantum channel will be separated by at least twice the classical channel spacing, i.e. 200 GHz (1.6 nm), so the LCXT from the next nearest classical channel into the quantum channel will be lowered $t_{i,n-a}$. For the assumed WDM 4[th] order Butterworth filtering with $t_{i,a}$ ~55 dB and $t_{i,n-a}$ ~82 dB (values that correspond to $t_{i,a}$ ~59 dB and $t_{i,n-a}$ ~82 dB used in [8]), the LCXT contribution from the next distant classical channel (removed by 300 GHz) is estimated as ~100 dB, which we shall neglect in our analysis. In principle, LCXT contributions from all the classical channels to the quantum channel can be accounted for if the properties of the classical WDM channel filters are known. For example, if Butterworth filtering is used and leads to 82 dB crosstalk for 200 GHz spacing, then it will lead to ~91-93 dB crosstalk for 250 GHz spacing, and ~103 dB crosstalk for 300 GHz spacing. That means we can neglect the contribution of more remote classical channel to the quantum channel LCXT. For 100 GHz and 50 GHz spacing such Butterworth filtering would lead to ~55 dB and ~32 dB crosstalk, respectively.

As will be discussed below (Section III.B), we assume that the choice of the frequency difference between any two co-propagating channels is such that no four wave mixing (FWM) frequency product is generated within the quantum channel passband; this helps to mitigate the direct impact of FWM.

Similar to the theoretical treatment in Ref. [8] our secret key rate estimation is optimistic, since we ignore any interruptions of the key exchange during key distillation and fiber length measurements. This approximation is more severe at higher key rates that typically occur for short fiber lengths.

### C. Table of used parameters

Table 1 summarizes all the parameters and the range of their numerical values used in our study.

**Table 1**: Model parameters

| Variable | Symbol | Typical value used | Units |
| --- | --- | --- | --- |
| fiber attenuation | $\alpha$[dB/km] | 0.16, 0.185, 0.195, 0.21, 0.3 | dB/km |
| fiber length | $L$ | 1 – 300 | km |
| fiber transmission | $t_F$ | $\exp(-\alpha L)$ | - |
| fiber dispersion | $D$ | 0.1, 0.16, 4.25, 20.35 | ps/nm/km |
| fiber dispersion slope | $S$ | 0.06 | ps/nm$^2$/km |
| Alice storage line length | $l_A$ | 10 | km |
| coefficient capturing the reduction of detection rate possible duty cycle imposed by the detection protocol synchronization requirements | $\eta_{\text{duty}}$ | $\sim \mathcal{O}(1)$ or Eq. (IV.5) | - |
| loss of receiver internal components | $t_B$[dB] | 2.65 | dB |
| WDM insertion loss (takes into account optical filtering in the receiver) – "increases the cross-talk by the same amount" | $t_{IL}$[dB] | 1.95 | dB |
| additional insertion loss of fiber Bragg gratings (FBG) | $t_{IL,FBG}$[dB] | 2 (when $\Delta\lambda_{FBG}$ is used) | dB |
| isolation of non-adjacent channels | $t_{i,n-a}$[dB] | 82 | dB |
| isolation of adjacent channels | $t_{i,a}$[dB] | 59 | dB |
| effective Raman cross-section (per km fiber length and nm bandwidth) | $\rho(\lambda)$ | $2 - 2.6 \times 10^{-9}$ | (km.nm)$^{-1}$ |
| quantum receiver bandwidth | $\Delta\lambda$ | 0.6 – 0.8 | nm |
| fiber Bragg grating bandwidth | $\Delta\lambda_{FBG}$ | 0.045 | nm |
| channel spacing | $\Delta\lambda_{ch}$ | 0.8 | nm |
| (classical) signal bit rate (permanent AES-256 transmission rate; data in one stream) | $f_{AB}$ | 1 – 200 | Gb/s |
| maximum secret key refresh period keeping security level the same (or fixed – in square brackets) | $T_{AES,max}$ | $60/(f_{AB}/1\text{ GHz})$ [or 60] | s |
| minimum secret key rate required for $N$ channel AES-256 encryption updated once in 60 s keeping security level the same (or keeping the refresh rate the same – in square brackets) | $f_{AB,min}$ | $(f_{AB}/1\text{ GHz}) \times N \times 256/60 \sim 4.27N \times (f_{AB}/1\text{ GHz})$ [or $N \times 256/60 \sim 4.27N$] | b/s |
| (classical) signal receiver sensitivity (depends on $f_{AB}$) | $R_x$ | from -28 to -50 | dBm |
| numbers of forward and backward classical channels | $N_f$ | 2 (1 for distillation, 1 for data encryption) | - |
| | $N_b$ | 2 (1 for distillation, 1 for data encryption) | - |
| classical channel power output from the fiber | $P_{out}$ | $R_x$[dBm] + $t_{IL}$[dB] + $t_{IL,FBG}$[dB] [Eq. (III.9)] | dBm |
| photon energy | $E_{photon}$ | $1.278818 \times 10^{-19}$ | J |
| CASCADE error correction protocol QBER distillation limit | $QBER_{thr}$ | 0.09 | - |
| CASCADE error correction algorithm correction to the number of discarded bits | $\eta_{ec}$ | 6/5 | - |
| number of APD quantum detectors | $N_d$ | 2 | - |
| Bob's "quantum" pulse generation rate | $f_{rep}$ | 5 – 20000 | MHz |
| Bob's "quantum" pulse generation period | $T = 1/f_{rep}$ | $50 - 200 \times 10^3$ | ps |
| dead time of an APD quantum detector | $\tau_{dead}$ | $2T$ or larger<br>0.002 – 10 for APD<br>0 or 0.3 for SNSPD | μs |
| quantum detector gate duration time | $\Delta t_{gate}$ | 1.0; ~$1/(2f_{rep})$ | ns |
| quantum detection efficiency | $\eta$ | APD:<br>0.07-0.19<br>($p'_{dc} \sim 5 - 8 \times 10^{-6}$ ns$^{-1}$)<br>SNSPD:<br>0.014 ($p'_{dc} \sim 50$ s$^{-1}$) | - |
| factor depending on QKD protocol | $\beta$ | $\beta_{BB84} = 1$;<br>$\beta_{SARG} = (2-V)/2$<br>$\beta_{COW} = 1$ | - |
| average number of photons per quantum channel pulse | $\mu$ | numerically optimized<br>[or, e.g., $\mu_{BB84} = t_F$;<br>$\mu_{COW} = 0.5$;<br>$\mu_{SARG} = 2(t_F)^{1/2}$] | - |
| signal detection probability | $p_\mu$ | $\mu \cdot t_F \cdot t_{IL} \cdot t_{IL,FBG} \cdot t_B \cdot \eta_{ISI}$<br>[Eq. (IV.2)] | - |
| dark count probability rate | $p'_{dc}$ | $5 - 8 \times 10^{-6}$/APD<br>$1 - 100 \times 10^{-9}$/SNSPD | ns$^{-1}$ |
| dark count probability | $p_{dc}$ | $p'_{dc} \cdot \Delta t_{gate}$ | - |
| Raman noise detection probability | $p_{ram}$ | Eqs. (III.6) | - |
| crosstalk photon detection probability rate | $p'_{LCXT}$ | Eqs. (III.8) | ns$^{-1}$ |
| crosstalk photon detection probability | $p_{LCXT}$ | $p'_{LCXT} \cdot \Delta t_{gate}$ | - |
| inter-symbol interference (ISI) detection probability due to chromatic dispersion | $p_{ISI}$ | Eq. (III.19) | - |
| loss of photons due to part of the pulse falling outside of quantum detector gate duration time | $t_{ISI}$ | Eq. (III.20)<br>~$\mathcal{O}(1)$ | - |
| pulse overlap with neighboring time gates due to ISI | $f_{err}^{(ISI)}$ | 0.001 | - |
| coefficient capturing the reduction of detection rate due to APD quantum detector dead time $\tau_{dead}$ | $\eta_{dead}$ | Eq. (IV.4) | - |
| after-pulse probability to the total detection probability fraction | $\rho_{AP}$ | APD: 0.008<br>SNSPD: 0 | - |
| after-pulse detection probability | $p_{AP}$ | Eq. (IV.3) | - |
| fringe visibility | $V$ | 0.9 – 0.998 | - |
| input quantum signal pulse duration | $\tau_{FWHM,0}$ | 10 – 50 | ps |
| quantum signal pulse spectral width | $\Delta \nu_{FWHM}$ | bandwidth limited value (Gaussian)<br>$\Delta \nu_{FWHM} = \dfrac{2 \ln 2}{\pi \cdot \tau_{FWHM,0}}$ | GHz |
| input quantum signal pulse chirp | $C$ | 0 | - |

## III. Classical/quantum WDM impairments

In order to treat the quantum channel impairments that the presence of classical channels brings along, we need to track the power in the classical channels: The output power of a classical channel, just before the detection, needs to be higher than a certain value to guarantee the required BER performance that is characterized by the detector sensitivity $R_x$; the maximum optical power in a classical channel is limited by fiber nonlinear impairments. Hence, classical signal detectors with larger sensitivity help mitigate the nonlinear impairments as do, of course, fibers with smaller nonlinearity. Such fibers are currently under study, e.g. [24],[25], but are not a subject of this study.

The quantum signal is impaired by loss, chromatic dispersion and Raman noise. When classical channels are present in the same fiber, there is an additional spurious photon noise, such as SpRS and FWM that are a consequence of fiber nonlinearity, and LCXT. Our analysis of the latter three factors will heavily rely on the approach given in Ref. [8]. In addition, the effects of fiber properties central to the sole QKD channel, the attenuation and chromatic dispersion, are also discussed.

## A. Spontaneous Raman scattering

The fundamental quantity characterizing SpRS is the effective Raman scattering cross-section $\rho(\lambda)$, measured as a function of the wavelength $\lambda$. The effective Raman scattering cross-section is intimately related to the commonly measured fiber characteristics, Raman gain [26]. Measured Raman gains were used to calculate $\rho(\lambda)$ of several types of single-mode fibers with different attenuations from the Corning Incorporated© portfolio (Ge-doped core, $SiO_2$-core, different index profiles, etc.). The results are shown in Fig. 1. Due to insufficient resolution, this procedure did not capture the region of small detuning from the continuous wave (cw) pump at 1495 nm wavelength. Nevertheless, the plots suggest small dependence of the effective Raman scattering cross-section on single-mode fiber type, with minimum SpRS impact less than $1.5\times10^{-9}$ $(km.sr.nm)^{-1}$ within a few nm detuning from the Raman pump. Also, the anti-Stokes side of the signal (lower than the Raman pump wavelengths) seems to be more favorable for the relative placement of the quantum channel(s).

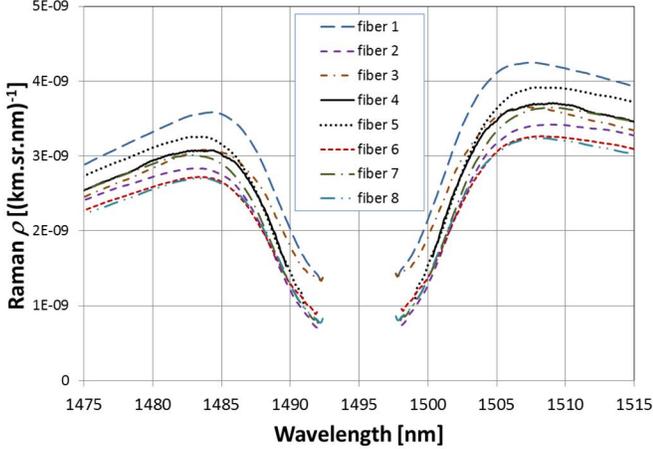

**Figure 1.** Impact of the SpRS for several single-mode fiber types. Laser source, i.e. the classical channel(s), in these simulations is at ~1495 nm; the SpRS is expected to be slightly smaller at laser sources at telecom wavelengths (~1550 nm) where the Raman gain is slightly smaller. Measured gain Raman spectra generated by a cw pump passing through different lengths of single-mode fiber were used to calculate Raman scattering cross-section $\rho(\lambda)$. (Note the scale does not show the peak power at the center frequency of the cw pump.)

The Raman scattered power emerging from the fiber input, $P_{ram,b}$, and fiber output, $P_{ram,f}$, can be calculated as

$$P_{ram,b} = N_b \cdot P_{out} \cdot \frac{\sinh(\alpha L)}{\alpha} \cdot \rho(\lambda) \cdot \Delta\lambda \qquad (III.1)$$

$$P_{ram,f} = N_f \cdot P_{out} \cdot L \cdot \rho(\lambda) \cdot \Delta\lambda \qquad (III.2)$$

where index $b$ stands for "backward," $f$ for "forward", $N_b$ and $N_f$ denote the number of backward and forward classical channels, $\Delta\lambda$ denotes quantum receiver bandwidth, and the optical power at the fiber output is

$$P_{out} = P_{in} \cdot e^{-\alpha L}, \qquad (III.3)$$

with $P_{in}$ denoting the power of the optical signal at the fiber input.

Using the Raman scatter powers the SpRS detection probabilities, $p_{ram,b}$ and $p_{ram,f}$, can be calculated using

$$p_{ram,f(b)} = \frac{P_{ram,f(b)}}{E_{photon}} \cdot \eta \cdot \Delta t_{gate}, \qquad (III.4)$$

where

$$E_{photon} = \frac{hc}{\lambda}. \qquad (III.5)$$

The total SpRS detection probability is given by

$$p_{ram} = p_{ram,f} + p_{ram,b}. \qquad (III.6)$$

## B. Four-wave mixing

FWM is a nonlinear process, in which the material and source photons interact to generate additional photons at new frequencies while preserving the energy-momentum conservation – no real excitation of the medium takes place. If the classical and quantum channel configuration is chosen poorly, stimulated FWM process can lead to the generation of photons at frequencies in the quantum channel [27]. Since the FWM effect on the quantum channel can be minimized by a proper choice of the classical channel separation, phase-matching conditions, and polarization, we assume that the classical and quantum channel configuration is chosen such that it prevents efficient FWM generation in the quantum channel band in standard single-mode fibers, dispersion shifted fibers and nonzero dispersion shifted fibers. In addition, the spontaneous FWM process contributes to the noise in the quantum channel band, but its effect is not a considerable contribution when compared to SpRS [8].

## C. Channel crosstalk

LCXT can be a significant source of noise in the quantum channel if the much more powerful classical channels are not sufficiently isolated. Acceptable isolation is a function of the classical signal receiver sensitivity $R_x[dBm]$ and the quality of the WDM filtering.

Knowing the energy of the photons, the classical signal receiver sensitivity can be converted into the minimum number of photons per nanosecond, $n_d$, that need to reach the detector to guarantee BER $< 10^{-12}$:

$$n_d\left[ns^{-1}\right] = \frac{10^{R_x[dBm]/10}}{E_{photon}[J]} \cdot 10^{-12}. \qquad (III.7)$$

This number of photons detected by APD is attenuated by the quantum detector efficiency $\eta$ and by the channel isolation $t_i$[dB], resulting in the detection probability rate

$$p'_{LCXT} = \eta \cdot n_d \left[ ns^{-1} \right] \cdot 10^{-t_i[dB]/10}, \quad (III.8)$$

which should be small compared to the intrinsic noise of a single quantum detector characterized by the dark count probability $p_{dc}$. In our simulations we distinguish between non-adjacent and adjacent channel isolations denoted as $t_{i,n-a}$[dB] and $t_{i,a}$[dB], respectively.

From the discussion of SpRS, FWM, and LCXT it is apparent that it is desirable to keep the total optical power propagating through the fiber as low as possible. It is really a restriction on the optical power in the classical channels which overwhelms any power propagating in the quantum channel. To detect a signal in a classical channel reliably and with small BER, however, a minimum power is required at the receiver of a classical channel, denoted as $R_x$. Since the optical power in a classical channel at the end of a fiber, $P_{out}$, is further attenuated by the presence of ("de-multiplexing") WDM and possible insertion loss of additional filtering using, e.g., fiber Bragg gratings (FBG, see Fig. 2) we can write:

$$P_{out}[dBm] = R_x[dBm] + t_{IL}[dB] + t_{IL,FBG}[dB]. \quad (III.9)$$

In budgeting the benefits of spectral filtering we note that there is a relationship between the amount of channel isolation and total IL, $t_{IL} + t_{IL,FBG}$: the larger the isolation the larger the total IL penalty and *vice versa*.

The insertion loss of the WDM before entering the fiber ("multiplexing" WDM) can be easily compensated by an increase of the source power for both classical and quantum channels. On the other hand, the total IL of the WDM and additional filtering is always present for both classical and quantum channels after the signals leave the fiber (while "de-multiplexing" the signals) and before they enter the corresponding detection scheme. The quantum signal, moreover, experiences the internal components loss $t_B$ on the quantum receiver side ("Bob"). This internal components loss $t_B$ equally attenuates all light impinging on the quantum detector, including the contribution from LXCT, SpRS, and FWM within the spectral width of the quantum channel.

### D. Fiber loss

Fiber loss has an important influence on QKD link efficiency in several ways. First, the larger the fiber loss the smaller the effective quantum bit rate [8]. Then, the larger the fiber losses the larger the QBER due to decreasing signal to quantum detector noise ratio. Also increasing fiber loss leads to a decrease of the QKD system robustness to photon number splitting (PNS) attacks [28]. It is clear that a fiber with the smallest possible loss is necessary and required for efficient QKD.

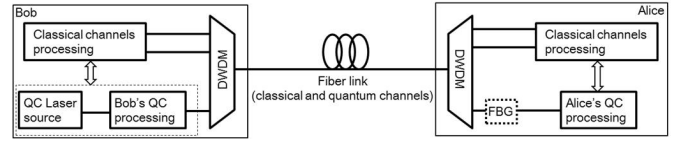

**Figure 2.** Basic setup for classical and quantum communication over the same fiber. Bob's QC processing involves $t_B$ loss for the quantum communication signal transmitted by Alice. This signal also suffers $t_{IL}$ from Bob's WDM. A classical signal leaving "classical channel processing" experiences $t_{IL}$ loss upon entry to WDM that can be compensated by an increase of the laser source output to guarantee required optical power at the input of the fiber $P_{in}$. The classical signal then propagates through the fiber and loses power. The power at the output of the fiber, $P_{out}$, is then further attenuated by the other WDM resulting in additional $t_{IL}$ loss, possibly additional IL of fiber Bragg gratings (FBG). The power after the second WDM should be compared to the receiver sensitivity. A quantum signal sent by Alice to Bob is attenuated to achieve $\mu$ photons per pulse at the input into the fiber, which means, accounting for the WDM loss $t_{IL}$ at Alice's side. The quantum signal then propagates through the fiber and loses power, then experiences $t_{IL}$ loss at Bob's WDM, $t_B$ loss at Bob's QC processing, and, possibly, additional IL of fiber Bragg gratings , before being detected with quantum detector efficiency $\eta$.

We capture the fiber loss, characterized by attenuation coefficient $\alpha$, as the fiber transmission, $t_F$, over a distance $L$, defined as

$$t_F \equiv e^{-\alpha L}. \quad (III.10)$$

### E. Chromatic dispersion

There is a considerable interest in increasing the quantum bit rate from 1 Mb/s to 1 Gb/s, 10 Gb/s, or higher [6],[7],[8],[9],[10],[11],[12]. In long QKD links based on optical fibers this may lead to signal impairments caused by chromatic dispersion.

To estimate the chromatic dispersion impact on QKD link performance, we assume an input quantum pulse with Gaussian-shaped input intensity (see Appendix B)

$$I(z=0,t) = |E_0|^2 \exp\left[-4\ln 2 \left(\frac{t}{\tau_{FWHM,0}}\right)^2\right]. \quad (III.11)$$

For simplicity, we assume zero initial chirp ($C = 0$) in the following. $\tau_{FWHM,0}$ denotes the initial full-width half-max (FWHM) pulse duration at the input. The pulse's envelope retains its Gaussian shape when propagating a distance $L$ through a linear medium; its pulse duration will however be altered due to chromatic dispersion:

$$I(z=L,t) \approx \exp\left[-4\ln 2 \left(\frac{t}{\tau_{FWHM,L}}\right)^2\right], \quad (III.12)$$

with $\tau_{FWHM,L}$ denoting the pulse's intensity FWHM which can be given in terms of the initial pulse duration $\tau_{FWHM,0}$, fiber link distance $L$, and the dispersion parameter $\beta_2 = (\lambda^3/2\pi c^2)(d^2n/d\lambda^2)$, conventionally given in ps$^2$/km, as

$$\frac{\tau_{FWHM,L}}{\tau_{FWHM,0}} \equiv \frac{\tau_{FWHM}(z=L)}{\tau_{FWHM,0}} = \sqrt{1 + \left(\frac{4\ln 2 |\beta_2| L}{\tau_{FWHM,0}^2}\right)^2} \quad \text{(III.13)}$$
$$\equiv \sqrt{1 + \left(\frac{L}{L_D}\right)^2}$$

Here $L_D$ designates the characteristic dispersion length. In the definition of $\beta_2$, $c$ denotes the speed of light in vacuum in nm/ps and $n$ the effective index of refraction of the propagating fiber mode. The dispersion parameter $\beta_2$ is related to the commonly used fiber dispersion parameter $D$ given in ps/nm/km by [29]

$$D = \frac{2\pi c}{\lambda^2} |\beta_2|. \quad \text{(III.14)}$$

We then have the following expressions for the characteristic dispersion length:

$$L_D \equiv \frac{\tau_0^2}{|\beta_2|} = \frac{\tau_{FWHM,0}^2}{4\ln 2 |\beta_2|}$$
$$= 2\pi \frac{c \cdot \tau_0^2}{\lambda^2 |D|} = \frac{\pi}{2\ln 2} \frac{c \cdot \tau_{FWHM,0}^2}{\lambda^2 |D|} \quad \text{(III.15)}$$

There are two effects caused by the gradual increase of a quantum signal pulse duration that we shall discuss here. Inter-symbol interference (ISI), which accounts for the effect of photons in the pulse falling into the gate time slot allocated to the neighboring bit and causing errors, and lowering the average number of photons in the pulse being detected within its gate time slot. Thus, the average number of photons per quantum channel pulse sent into the fiber, $\mu$, will be diminished by the (multiplicative) amount

$$t_{ISI} \equiv \int_{-\Delta t_{gate}/2}^{\Delta t_{gate}/2} dt\, I(L,t). \quad \text{(III.16)}$$

Normalized input and output Gaussian pulse intensity shapes

$$I(z,t) = \frac{1}{\sqrt{\pi}} \frac{2\sqrt{\ln 2}}{\tau_{FWHM,z}} \exp\left[-4\ln 2\left(\frac{t}{\tau_{FWHM,z}}\right)^2\right]; \quad z = 0, L, \quad \text{(III.17)}$$

are illustrated in Fig. 3. The error probability due to ISI impending from one of the neighboring time bits (only nearest neighbor bit slots are considered) will be proportional to a fraction of the elongated pulse that overlaps with the gate, given by

$$f_{err}^{(ISI)} = \int_{T-\Delta t_{gate}/2}^{T+\Delta t_{gate}/2} dt\, I(L,t)$$
$$= \frac{2\sqrt{\ln 2}}{\sqrt{\pi}} \frac{1}{\tau_{FWHM,L}} \cdot \int_{T-\Delta t_{gate}/2}^{T+\Delta t_{gate}/2} dt \exp\left[-4\ln 2\left(\frac{t}{\tau_{FWHM,L}}\right)^2\right],$$
$$= \frac{1}{2}\left[erfc\left(\frac{\sqrt{\ln 2}}{\tau_{FWHM,L}}(2T - \Delta t_{gate})\right) - erfc\left(\frac{\sqrt{\ln 2}}{\tau_{FWHM,L}}(2T + \Delta t_{gate})\right)\right]$$
(III.18)

where we used the complementary error function, denoted $erfc$ and defined as $erfc(x) \equiv (2/\sqrt{\pi})\int_x^\infty dt \exp(-t^2)$. We introduce the ISI error detection probability due to chromatic dispersion as

$$p_{ISI} = 2 \cdot f_{err}^{(ISI)} \cdot \mu \cdot t_F \cdot t_{IL} \cdot t_{IL,FBG} \cdot t_B \cdot \eta, \quad \text{(III.19)}$$

where we accounted for contributions from two neighboring time bits and the losses the quantum signal experiences after being coupled to the fiber.

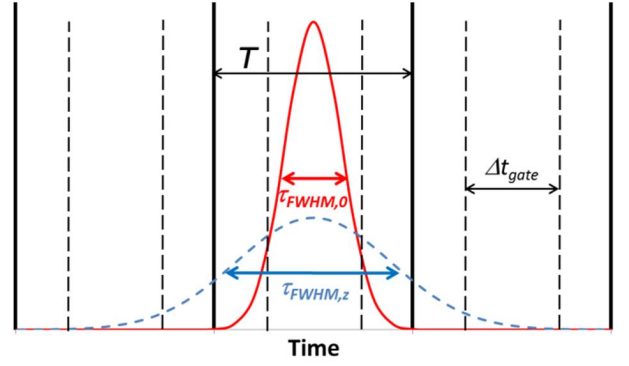

**Figure 3.** Input (full red line) and output (dashed blue line) Gaussian pulses. $T$ – quantum bit period, $\tau_{FWHM,0}$ – width of input pulse, $\tau_{FWHM,L}$ – width of broadened output pulse, $\Delta t_{gate}$ – detection time window.

The reduction in the average number of photons per quantum channel pulse sent into the fiber can be obtained by substituting Eq. (III.17) into the expression given by Eq. (III.16):

$$t_{ISI} = \frac{1}{2}\left[erfc\left(-\Delta t_{gate}\frac{\sqrt{\ln 2}}{\tau_{FWHM,L}}\right) - erfc\left(\Delta t_{gate}\frac{\sqrt{\ln 2}}{\tau_{FWHM,L}}\right)\right]. \quad \text{(III.20)}$$

From the above considerations, it is clear that the relation between the initial pulse duration $\tau_{FWHM,0}$ and the detection time window $\Delta t_{gate}$ can be treated as an additional optimization parameter. In this study, we do not perform this optimization, but fix the ratio at a constant value.

In general, quantum links employ pulses with extremely low photon counts. The lack of photons per signal pulse prevents usage of digital signal processing techniques common in optical telecommunications to electronically compensate chromatic and, partially, polarization mode dispersion introduced by the fiber medium. Instead, analog dispersion compensation needs to be implemented, if warranted, even at the cost of an implementation penalty or increased loss.

There are, perhaps, three methods of analog dispersion compensation in optical fiber links [30]. The first is applying dispersion compensation fibers (DCF) after the signal propagation (post-compensation). The typical parameters of DCF are 0.6 dB/km attenuation and -100 ps/nm/km chromatic dispersion [31]. The length of DCF that compensates 300 km of Corning® Vascade® EX2000 optical fiber with $D_{EX2000}$ = 20.35 ps/nm/km dispersion is calculated as

$$L_{DCF} = \frac{L_{EX2000} \cdot D_{EX2000}}{|D_{DCF}|} \approx 60 \text{ km.} \quad \text{(III.21)}$$

We conclude that DCF attenuation of 36 dB leads to problems in signal power control. An alternative method of chromatic dispersion post-compensation is applying an FBG or chirped Bragg grating [30]. The advantage of this method is a much smaller attenuation [30],[32]. In paper [32] an insertion loss < 0.5 dB is reported. The third way to battle the effect of chromatic dispersion on a pulse is to engineer the quantum pulse chirp in such a way that the shortest pulse arrives within the desired time; in other words, to pre-compensate the chromatic dispersion effect, e.g. using DCF, FBG, or chirped Bragg gratings. The attenuation of DCF does not play a role in this approach, since it is applied before the quantum signal is attenuated to the desired average photon number in a pulse, $\mu$. Even though we do not investigate this option in detail here, the theory presented in the current section (III.E), Section V.B, and Appendix B, is sufficient to capture the main impact of DCF pre-compensation on QBER and $R_{sec}$. In this work we consider only dispersion uncompensated links without pre-compensation.

### F. Polarization mode dispersion

In order to complete the discussion of the impact of fiber impairments on quantum communication that uses short pulses, we briefly touch on the topic of polarization mode dispersion (PMD). Since most modern fibers have PMD under control with values of typically below 0.05 ps/sqrt(km) [33], the total pulse spreading occurring over 400 km fiber span is less than 1 ps. So unless the pulses used to send the quantum information, and the associated gate times, are in the sub-picosecond regime, PMD in QKD links using such fibers may be neglected.

Besides the PMD induced pulse broadening, fiber PMD also leads to a polarization state instability during the quantum pulse propagation. Such polarization rotation or change is an important issue for some QKD protocols like BB84, but do not play significant role for other ones, such as COW. In this work we assume that the fiber quality is such that any PMD induced polarization state changes are not detrimental to any QKD protocol we study.

## IV. Classical and quantum detection

The performance of a classical telecommunication channel can be characterized in several ways that are not independent such as BER, reach or capacity, i.e., how much information can be transmitted through the channel in a unit of time. Similarly, the metrics of interest for a quantum channel are its QBER, its reach, and secret key rate. We show that the use of CMF improves the performance attributes of the QKD channel, since it enables a reduction of the SpRS, thus increasing the length or the capacity of QKD link. This benefit increases in conjunction with low loss fibers.

### A. Classical channel – advanced modulation formats and FEC

As we discussed above, when assuming QKD and classical channels propagating in a single fiber core, the number of quantum and classical channels is limited by SpRS. Due to the dark count inherent in the detection of QKD signals (i.e. only a few photons) one can allow only a certain level of SpRS noise before encountering its negative effects. To limit the impact of SpRS on the quantum channel, the number of quantum and classical channels is restricted and the propagation length depends on the receiver sensitivity, which is a function of modulation format and use of FEC techniques for the classical channel(s).

The receiver sensitivity of several modern CMF used in the transmission of a classical signal are shown in Fig. 4 for 10 Gbaud symbol rate. The details of the calculations are deferred to Appendix A. The parameters of the model, including the 10 Gbaud symbol rate, were chosen such that the results in left and right panes of Fig. 4 are in agreement with those found in Ref. [34], Fig. 2.38, for both theoretical (no implementation penalty) and measured BERs, respectively. In order to obtain receiver sensitivity curves that agree with the experimentally measured curves reported in Ref. [34], Fig. 2.38(a), we also considered the effect of an implementation penalty in a model described in App. B.

Notice that current FEC protocols require an input BER $\sim 10^{-3.0}$ for hard-decision (HD) and $\sim 10^{-2.4}$ for soft-decision (SD) algorithms, in order to output a BER $< 10^{-12}$. In particular, we estimate on the basis of the measured results in Fig. 4 that a BPSK receiver has a sensitivity of -48 dBm when HD is used and -50 dBm when SD is used. Similarly, QPSK receiver has sensitivity -45 dBm when HD is used and -47 dBm when SD is used. Note, that the receiver sensitivity is equal to the received power necessary to achieve desired performance; more precisely, the power at the receiver, which contains several individual classical signal detectors, in both polarizations.

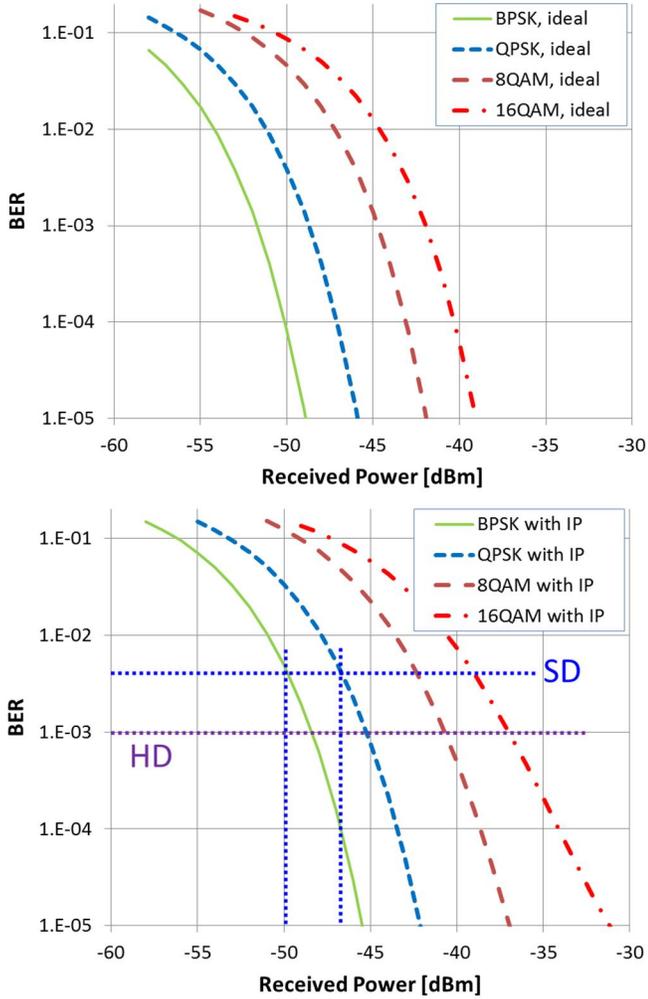

**Figure 4.** Receiver sensitivity of a few coherent transmission modulation formats for the classical channel @10 Gbaud. Left panel: Ideal modulation format, no implementation penalty (IP) assumed. Right panel: To mimic the effects of realistic implementation and hidden transmitter noise, IP is added to the model. This model agrees well with the experimental data in Ref. [34]. SD – soft decision FEC threshold BER, $\sim 10^{-2.4}$, HD – hard decision FEC threshold BER, $\sim 10^{-3.0}$. According to the data in the right panel, -50dBm output signal power is sufficient to reach SD FEC threshold for PM-BPSK modulation format, -47 dBm for PM-QPSK, and -38 dBm for PM-16QAM, approximately.

We want to stress that the results presented in Fig. 4 were obtained for classical communication systems employing 10 Gbaud symbol rate (corresponding to, e.g. 40 Gb/s PM-QPSK or 20 Gb/s PM-BPSK) that is associated with a particular electrical shot noise (-58.5 dBm; see Appendix A). Modern communication systems use even higher symbol rates with limitations coming from the available speed of the electronics. For 32 Gbaud symbol rate, assuming the same implementation penalty, the receiver sensitivity increases by additional 10 $\log_{10}(32/10) \sim 5$ dB with respect to the receiver sensitivity observed at 10 Gbaud symbol rates.

Finally, we want to point out that the receiver characteristics quote here may be associated with lab-grade receivers, which might not be commercially available yet, and may represent the current state of the art.

### B. Quantum channel detection

There are two important characteristics related to the detection of the quantum signal: the rate at which the secret key is delivered to the recipient and the number of bit errors in the sifted key called the QBER, which determines the quality of the quantum signal.

The secret key rate calculation starts with the evaluation of the raw detection rate $R_{raw}$ delivered by the detectors due to quantum signals, quantum detector dark counts, after-pulses and additional noise [8],

$$R_{raw} = \left( p_\mu + N_d p_{dc} + p_{AP} + p_{ram} + p_{LCXT} + p_{ISI} \right) f_{rep} \eta_{duty} \eta_{dead} .$$
(IV.1)

Here, $f_{rep}$ is the (quantum signal) pulse repetition frequency of the system and the quantities $p_x$ signify detection probabilities per quantum detector gate. In particular, $p_\mu$ denotes signal, $p_{dc}$ dark count, $p_{AP}$ after-pulse, $p_{ram}$ Raman photon, $p_{LCXT}$ crosstalk photon detection probabilities, and $p_{ISI}$ the ISI error detection probability due to chromatic dispersion. $N_d$ is the number of APD quantum detectors – in our simulations always equals to 2 for COW and BB84 protocols (equals to the dimension of the bases used in the QKD protocol).

The quantum signal detection probability is given by

$$p_\mu = \mu \cdot t_F \cdot t_{IL} \cdot t_{IL,FBG} \cdot t_B \cdot t_{ISI} \cdot \eta ,$$
(IV.2)

the after-pulse detection probability can be approximated by

$$p_{AP} \approx \rho_{AP} \cdot \left( p_\mu + N_d p_{dc} + p_{ram} + p_{LCXT} + p_{ISI} \right),$$
(IV.3)

where $\rho_{AP}$ denotes after-pulse probability to the total detection probability fraction. $p_{ram}$, $p_{LCXT}$, and $p_{ISI}$ are given by Eq. (III.6), Eq. (III.8), and (III.19), respectively. Finally, to account for the reduced detection rate due to a quantum detector dead time $\tau_{dead}$ applied after each detection, we use the coefficient $\eta_{dead}$:

$$\eta_{dead} = \left[ 1 + \tau_{dead} f_{rep} \left( p_\mu + N_d p_{dc} + p_{AP} + p_{ram} + p_{LCXT} + p_{ISI} \right) \right]^{-1},$$
(IV.4)

and the coefficient $\eta_{duty}$ to capture possible duty cycle imposed by the detection protocol synchronization requirements, e.g. like the "plug and play" from IDQ used in Ref. [8]:

$$\eta_{duty} = \frac{l_A}{L + l_A} .$$
(IV.5)

In general, we treat $\eta_{duty}$ as a parameter characterizing the effect of different possible synchronization schemes on the detection rate that is of order of 1, $\sim \mathcal{O}(1)$, but smaller or equal to 1.

The quantum detector dead time $\tau_{dead}$ can be estimated by utilizing the review [35]. There, we can find that the best detector performance has been achieved with self-differencing InGaAs single-photon avalanche diode (SPAD) operating at 240 K with efficiency $\eta = 10\%$ at 1550 nm wavelength and at the clock rate $f_{rep} = 1.036$ GHz. The smallest separation between the avalanches attained 2 ns, in accord with the theoretical limit for this type of detector that equals to 2 clock periods, $\tau_{dead} \sim 2T = 2/f_{rep}$. In this paper we treat $\tau_{dead}$ as a fitting parameter, studying its effect on the QKD system performance explicitly for both SPAD and superconducting nanowire single-photon detectors (SNSPD). Quantum clock rates up to 10 GHz have been reported [4],[6] using 15 ps pulses and SNSPDs with low dark count (~50 Hz) and a small timing jitter (~60ps). While these values of SNSPD parameters far outperform the SPAD, they come at a cost of much smaller quantum efficiency, as small as 1%, and operating temperature of a few Kelvins [4]. Furthermore, Ref. [35] provides us also with a hint for the smallest value of the quantum detector gate time being approximately half of the clock period, $\Delta t_{gate} = T/2 = 0.5/f_{rep}$.

A certain fraction of $R_{raw}$ is discarded and the "sifted" key rate is obtained as

$$R_{sift} = \frac{1}{2}\left(\beta p_\mu + N_d p_{dc} + p_{AP} + p_{ram} + p_{LCXT} + p_{ISI}\right) f_{rep} \eta_{duty} \eta_{dead}.$$

(IV.6)

The sifting algorithm depends on the QKD protocol, as depicted by the parameter $\beta$ that varies from QKD protocol to QKD protocol. Denoting $I_{AB}$ and $I_{AE}$ as the mutual information per bit between Alice and Bob, and between Alice and a potential eavesdropper, respectively, the secret key rate $R_{sec}$ after error correction and privacy amplification is estimated as (incoherent attacks [8])

$$R_{sec} = R_{sift}\left(I_{AB} - I_{AE}\right).$$

(IV.7)

The mutual information per bit between Alice and Bob is given by

$$I_{AB} = 1 - \eta_{ec} H(QBER);$$
$$H(p) \equiv -p \log_2 p - (1-p) \log_2(1-p)$$

(IV.8)

Here, $H(p)$ is the Shannon entropy for a given QBER $p$ that is related to the minimum fraction of bits lost due to error correction. CASCADE error correction algorithm penalty is captured by $\eta_{ec} = 6/5$, a correction to the number of discarded bits, since CASCADE cannot reach the theoretical Shannon limit and since a certain fraction of distilled secret bits is consumed for authentication.

$I_{AE}$ depends on the particulars of the algorithm that is used to combat Eve's attacks, e.g., for BB84 protocol one can write [8],[36]

$$I_{AE,BB84} = \frac{\left(1 - \frac{\mu}{2t_F}\right)\left[1 - H(P)\right] + \frac{\mu}{2t_F}}{1 + \frac{N_d p_{dc}}{\mu t_F \eta}};$$

$$P = \frac{1}{2} + \sqrt{d(1-d)}; \quad d = \frac{1-V}{2 - \frac{\mu}{t_F}}$$

(IV.9)

For COW we use [37],[18] to estimate $I_{AE}$ as

$$I_{AE,COW} = \mu(1-t_F) + (1-V)\frac{1+e^{-\mu t_F}}{2e^{-\mu t_F}},$$

(IV.10)

where the first term corresponds to individual beam splitting attacks and the second to intercept-resend attacks [37], when PNS attacks do not introduce errors. The estimate given by Eq. (IV.10) representing the COW protocol security against a large class of collective attacks assumes that Bob receives at most one photon per bit. Also, the above theoretical estimates ignore interruptions of the key exchange during key distillation and various experimental manipulations, such as fiber length measurements. Such influence is more significant for higher key rates, when Eqs. (IV.1)-(IV.10) overestimate the secret key rate, especially for short fiber lengths.

For fiber lengths more than 100 km, the fiber transmission $t_F$ becomes rather small and in order for Eq. (IV.9) to work $\mu$ has to be chosen appropriately small as well. The choice of $\mu$ needs to be such that the Eq. (IV.7) remains positive, i.e., $I_{AB} > I_{AE}$: We require a choice of $\mu$ such that $D \in (0,1)$ in Eq. (IV.9), at least. It has been shown that, within the region of validity of the Eqs. (IV.7)-(IV.9), $\mu = t_F$ is the analytically obtained optimum for $R_{sec}$ for fibers of moderate length (for which $N_d p_{dc} \ll \mu t_F \eta$) and the BB84 protocol [36]. For longer fiber lengths Eq. (IV.9) needs to be optimized numerically to preserve the restriction on $d$. The optimization shows that in the range of parameters studied in this paper, there is no practical difference between the optimized $\mu$ values and $\mu = t_F$ for BB84 protocol. Note that in the present work we optimize $\mu$ using Eq. (IV.9), hence we optimize for maximal $R_{sec}$, even at the expense of possible QBER deterioration. The levels of QBER at the optimal values of $\mu$ are typically low enough to be acceptable for the quantum FEC processing.

For completeness, we also quote the expression for $I_{AE,SARG}$ from Ref. [8]:

$$I_{AE,SARG} = I_{pns}(1) + \frac{1}{12}\frac{\mu^2}{t_F}e^{-\mu}\left(1 - I_{pns}(1)\right),$$

(IV.11)

with $I_{pns}(k) = 1 - H\left[0.5\left(1 + \sqrt{1 - 1/2^k}\right)\right]$ denoting the potential information gain of an eavesdropper due to PNS attacks on multi-photon pulses when $k$ photons are split and stored.

QBER, the number of errors present in the key obtained after the sifting, can be written, assuming that the visibilities in all the interferometric bases are the same, for BB84 (and SARG) as [8],[37]

$$QBER_{BB84} = \frac{1}{2} \frac{p_\mu(1-V) + N_d p_{dc} + p_{AP} + p_{ram} + p_{LCXT} + p_{ISI}}{\beta p_\mu + N_d p_{dc} + p_{AP} + p_{ram} + p_{LCXT} + p_{ISI}},$$

(IV.12)

and for COW protocol as [37]

$$QBER_{COW} = \frac{1}{2} \frac{N_d p_{dc} + p_{AP} + p_{ram} + p_{LCXT} + p_{ISI}}{\beta p_\mu + N_d p_{dc} + p_{AP} + p_{ram} + p_{LCXT} + p_{ISI}}.$$

(IV.13)

## V. Results

Most previous QKD experiments used multiple single-core fibers to transmit classical and quantum signals separately. Early efforts to multiplex classical and quantum channels onto a single optical link resulted in relatively short signal propagation distance (30-40 km) due to the detrimental effect of SpRS on the quantum channels. Suppressing SpRS would enable a longer quantum signal propagation distance. In this paper, we consider use of CMF and FEC as strategies to lower the impact of SpRS on the quantum optical link. Some basic results for each of the approaches are presented below.

We start with the evaluation of the impact of CMF and FEC on SpRS. Then we study the effect of CMF, FEC, (classical channel) bit rate, and the number of classical channels on the QBER and secret key rate.

### A. SpRS

The simulated impact of CMF and FEC on SpRS is shown in Fig. 5 as a function of propagation length. Low loss fiber (0.16 dB/km) was assumed. Since the SpRS is smaller at wavelengths below the pump, i.e. classical channel(s), the quantum channel(s) are likely to be transmitted at wavelengths lower that the classical channels. The maximum SpRS was assumed to be $2.6 \times 10^{-9}$ (km.nm)$^{-1}$, since some of Corning's existing single mode fibers can achieve such values or smaller for all wavelengths below the ~1550 nm classical channel wavelength. If only a few classical channels are considered, say up to 4 or 5, a value of $2.0 \times 10^{-9}$ (km.nm)$^{-1}$ per channel can be taken for SpRS, since it is achievable by existing fibers.

In our analysis, SpRS noise is compared to the intrinsic dark count of a typical APD and zero LXCT is assumed. The (classical channel) receiver sensitivity is based on the choice of CMF and FEC for 10 Gbaud symbol rate transmission systems (see Section IV.A and Fig. 4). The plots in Fig. 5 suggest that ~100 km of several WDM quantum channels could be propagated in the presence of one bi-directional classical channel modulated by 10 Gbaud symbol rate PM-

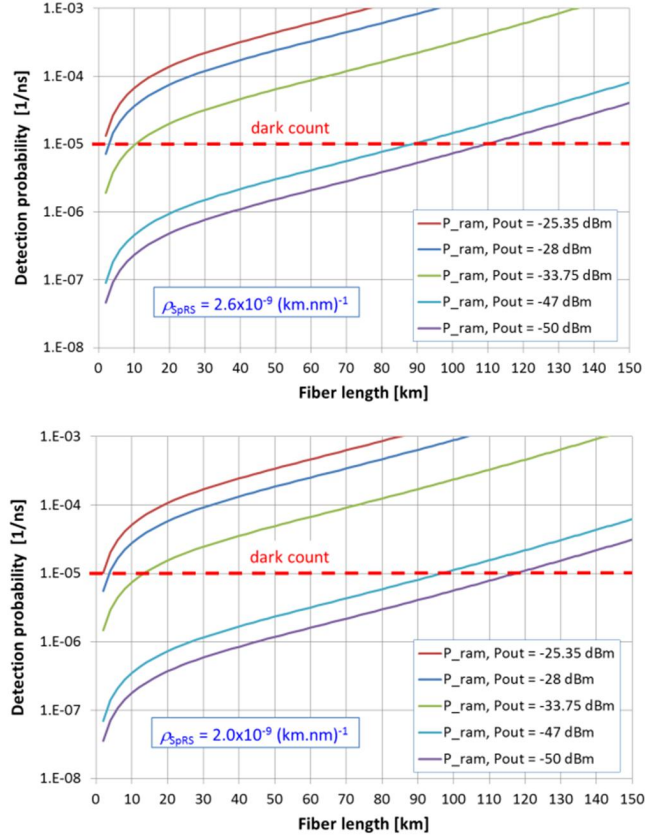

**Figure 5.** Comparison of detection probabilities of intrinsic quantum detector noise (dark counts and afterpulsing; red dashed line) and SpRS noise as a function of the optical power $P_{out}$ at the output of the fiber, 0.16 dB/km fiber loss, maximum SpRS of $2.6 \times 10^{-9}$ (km.nm)$^{-1}$ (top pane) and $2.0 \times 10^{-9}$ (km.nm)$^{-1}$ (bottom pane). Note that, the minimal receiver sensitivity needed to detect the signal is given by Eq. (III.9): $R_x = P_{out}$[dB] - $t_{IL}$ - $t_{IL,FBG}$. One bi-directional classical channel, two APDs with $p'_{dc} = 0.5 \times 10^{-5}$ ns$^{-1}$/APD, $\Delta t_{gate} = 1$ ns, $\eta = 0.07$, $\Delta \lambda = 0.8$ nm, and $p_{LXCT} = 0$ were assumed.

QPSK (or two bi-directional classical channels modulated by 10 Gbaud symbol rate PM-BPSK format, since data rate-wise one PM-QPSK channel transmits the same amount of information per unit time as two PM-BPSK channels and this factor of 2 erases the 3 dB benefit of the PM-BPSK channel in the receiver sensitivity) before the effect of SpRS would dominate the impact on the performance of quantum channels. Note that in the above estimate we neglected the total IL and assumed receiver sensitivities according to Ref. [34] and Fig. 4. We also note that typically four one-directional (two bi-directional) classical channels are required for sufficient transfer of quantum key distillation data and encrypted data in both directions between Alice and Bob [8].

Analysis of SpRS leads to several observations: (i) Use of more classical channels, according to our current understanding, would dramatically decrease the number of quantum channels in the WDM with acceptable performance or decrease the propagation length. (ii) The improved sensitivity of modern coherent-based classical signal detectors relaxes the requirements on the channel isolation provided by WDM filters. (iii) The number of quantum and classical channels and the propagation length also depends on the fiber loss; smaller loss allows us to increase the maximal distance in QKD link.

Throughout Section V we investigate, in particular, the 1 Gb/s QKD system described in Ref. [11] and its modifications from different points of view. The specific parameters we associate with this system are listed in the caption to Fig. 6. In Fig. 6 we show the results of evaluating the SpRS in the presence of two bi-directional classical channels in such a system. We finish the section with the discussion of 10 Gb/s QKD systems.

### B. Chromatic dispersion and QBER

There is an optimal initial pulse duration for which the relative broadening caused by chromatic dispersion is the smallest. That opens the possibility of pulse duration optimization for the QKD link. Before addressing this issue, we focus on estimating ISI for a given pulse duration sufficiently smaller than the quantum bit period for $\leq 500$ km length of fibers in current use. We use pulse durations that would lead to a sufficiently small impact of ISI after propagation over on the order of tens of picoseconds and more.

We shall fix $\tau_{FWHM,0} = 0.1T$ and $\Delta t_{gate} = T/2$ [11]. We can see from the expressions for QBER in Section IV.B, Eqs. (IV.12)-(IV.13), that we may neglect the influence of $p_{ISI}$ with respect to the quantum signal detection probability $p_\mu$, Eq. (IV.2), and other sources of detection errors in Eq. (IV.1), if $0.5 p_{ISI}/p_\mu = f_{err}^{(ISI)}/t_{ISI} \ll QBER$. This is a consequence of the requirement $QBER - QBER|_{p_{ISI} \equiv 0} \ll QBER|_{p_{ISI} \equiv 0} \approx QBER$ and the relative size of terms in the expressions for $QBER$, leading to the condition $QBER - QBER|_{p_{ISI} \equiv 0} \approx 0.5 p_{ISI}/p_\mu$. The constraint that at the fiber propagation length $L$ the ISI penalty can be neglected while keeping $QBER(L) \leq QBER_{thr}$, means that we can relax the above condition and demand $QBER - QBER|_{p_{ISI} \equiv 0} \ll QBER_{thr}$. The assumptions that the QBER threshold $QBER_{thr} = 0.09$ [8] and $t_{ISI}$ is of order of 1, $\sim \mathcal{O}(1)$, leads to the estimate $f_{err}^{(ISI)} \leq 0.001$. In this case we can write

$$f_{err}^{(ISI)} \approx \frac{1}{2} erfc\left[\frac{\sqrt{\ln 2}}{\tau_{FWHM,L}}\left(2T - \Delta t_{gate}\right)\right], \quad (V.1)$$

Substituting $\tau_{FWHM,L}$ from Eq. (III.13) into Eq. (V.1) we get:

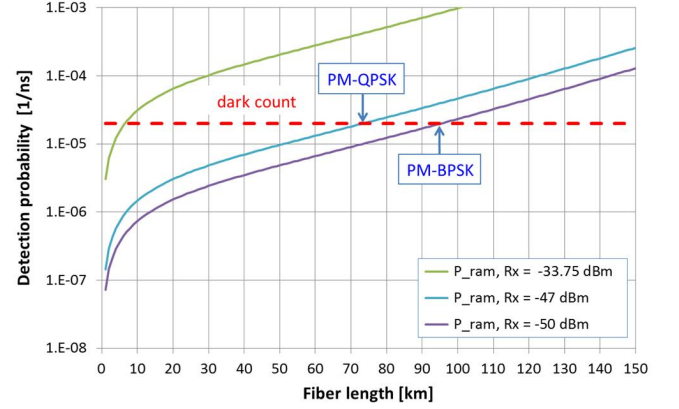

**Figure 6.** Comparison of detection probabilities of intrinsic quantum detector noise (dark counts and afterpulsing; red dashed line) and noise caused by SpRS in two bi-directional classical channels as a function of receiver sensitivity for fiber similar to Corning® Vascade® EX2000 optical fiber. $P_{out}[dBm] = R_x + t_{IL} + t_{IL,FBG}$.

| Fiber | Corning® Vascade® EX2000 optical fiber; 0.16 dB/km loss; two bi-directional 10 Gbaud classical channels |
|---|---|
| SpRS effective Raman cross-section | $2.6 \times 10^{-9}$ (km.nm)$^{-1}$ |
| Dark count | $1 \times 10^{-5}$ ns$^{-1}$/APD |
| $\Delta t_{gate}$ (at $f_{rep} = 1$ Gb/s) | 0.5 ns |
| $\eta$ | 0.19 |
| $\Delta \lambda$ | 0.6 nm |
| quantum channel isolation | 82 dB |
| $t_{IL} + t_{IL,FBG}$ | 1.95 dB |

$$f_{err}^{(ISI)} \approx \frac{1}{2} erfc\left[\frac{\sqrt{\ln 2}}{\tau_{FWHM,0} \cdot \sqrt{1 + \left(\frac{L}{L_D}\right)^2}}\left(\frac{2}{f_{rep}} - \Delta t_{gate}\right)\right]_{f_{rep} = v_{max}}, \quad (V.2)$$

with $L_D$ given in Eq. (III.15). Note that we also introduced the quantum bit rate $f_{rep} = 0.1/T$. Now, for example, we may find the maximal achievable quantum bit rate $f_{max} \equiv \max\{f_{rep}\}$ by solving equation (V.2) with the selected values $f_{err}^{(ISI)} = 0.001$, $\tau_{FWHM,0} = 0.1T = 0.1/f_{rep}$ and $\Delta t_{gate} = T/2 = 0.5/f_{rep}$. Assuming units of $D$ in ps/nm/km, $l$ in nm, $c$ in ps/nm, $L$ in km, and $f_{max}$ (or $f_{rep}$) in GHz, we obtain

$$0.001 = \frac{1}{2} erfc\left[ \frac{15\sqrt{\ln 2}}{\sqrt{1 + \left(\frac{2\ln 2 \cdot \lambda^2 \cdot |D| \cdot f_{max}^2 \cdot L}{10000 \cdot \pi \cdot c}\right)^2}} \right].$$

(V.3)

Using Eqs. (III.20), (III.13), and (III.15), the number of photons detectable within the gate time will be reduced to

$$t_{ISI} = \frac{1}{2}\left[ erfc\left( -\frac{5\sqrt{\ln 2}}{\sqrt{1 + \left(\frac{2\ln 2 \cdot \lambda^2 \cdot |D| \cdot f_{max}^2 \cdot L}{10000 \cdot \pi \cdot c}\right)^2}} \right) - erfc\left( \frac{5\sqrt{\ln 2}}{\sqrt{1 + \left(\frac{2\ln 2 \cdot \lambda^2 \cdot |D| \cdot f_{max}^2 \cdot L}{10000 \cdot \pi \cdot c}\right)^2}} \right) \right].$$

(V.4)

Since *erfc* in Eq. (V.3) attains a value of 0.002 at its approximate argument equal to 2.185124, we obtain, by comparing the argument of *erfc* functions in Eq. (V.4) with the argument of the *erfc* function in Eq. (V.3),

$$\begin{aligned} t_{ISI} &\approx 0.5\left[ erfc(-5/15 \cdot 2.185124) - erfc(5/15 \cdot 2.185124) \right] \\ &\approx 0.5\left[ erfc(-0.72837467) - erfc(0.72837467) \right] \approx 0.697 \end{aligned}.$$

That means, that for the choices of $f_{err}^{(ISI)}$, $\tau_{FWHM,0}$, and $\Delta t_{gate}$ made above, the mean number of photons is reduced by an approximate multiplicative factor 0.7, which does not affect the QBER and secure key rate significantly.

Solving Eq. (V.3) results in the $f_{max}(|D|,L)$ dependence illustrated in Fig. 7 below. The dispersion parameters of Corning® Vascade® EX2000 optical fiber, Corning® LEAF® optical fiber, low dispersion fiber (LDF), also known as dispersion shifted fiber (DSF), and standard single mode fiber (we refer to the family of standard single-mode fibers as SMF28e® optical fiber) are listed in Table 2. The 0.3 dB/km value for SMF28e® optical fiber loss was used in one instance to compare our modeling results with those experimentally achieved in Ref. [11]. Note that, parameters used in simulation are rough characteristic of the fiber family, not the specifications associated with the representative fibers, even though the modeled parameters do not necessarily agree with their specification for each representative fiber.

Considering 1 Gb/s QKD links, the estimates in Fig. 7 show that the impact of chromatic dispersion in current fibers is negligible. We conclude that fibers with $|D| < 25$ ps/nm/km, which is the majority of the deployed fibers and fibers in production, can support 1 Gb/s quantum bit rate QKD without significant influence of chromatic dispersion over distances $L \approx 500$ km or even larger. So far, 1 Gb/s system has been deployed over 45 km distance over an ITU-T G.652 compatible, single-mode fiber [11].

The impact of chromatic dispersion in 10 Gb/s QKD systems needs to be accounted for more carefully, if fibers with typical *D* are employed. The results in Fig. 7 suggest that a fiber with $|D| \leq 0.1$ ps/nm/km should be sufficient to support 10 Gb/s quantum bit rate QKD without significant influence of chromatic dispersion over substantial distances (assuming no chromatic dispersion pre-compensation).

**Table 2**: Fiber properties. (Parameters used in simulation are rough characteristic of the fiber family, not the specifications associated with the representative fibers. In the text, we refer to the fiber families by their representative fibers in this table, even though the modeled parameters do not necessarily agree with their specification for each representative fiber.)

| Fiber family label | Representative fiber | Dispersion @1550nm [ps/nm/km] | Dispersion slope @1550nm [ps/nm$^2$/km] | Modeled loss [dB/km] |
|---|---|---|---|---|
| $\langle 1 \rangle$ | Corning® Vascade® EX2000 optical fiber | 20.35 | 0.06 | 0.16 |
| $\langle 2 \rangle$ | Corning® LEAF® optical fiber | 4.25 | 0.085 | 0.185 |
| $\langle 3 \rangle$ | LDF/DSF | 0.1 | 0.085 | 0.185 |
| $\langle 4 \rangle$ | SMF28e® optical fiber | 17 | 0.06 | 0.21 (0.3) |

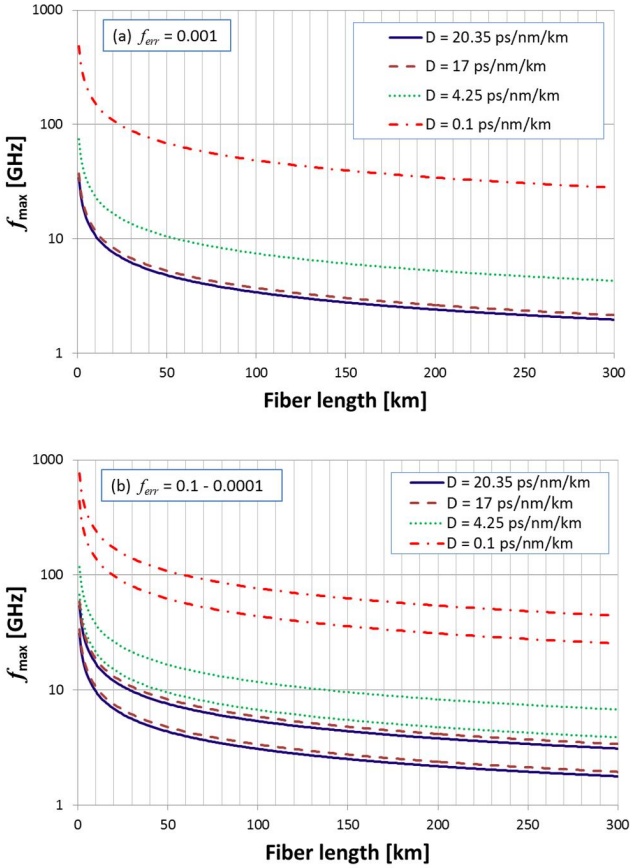

**Figure 7.** Maximum quantum bit rate $f_{max}$, found as the solution of Eq. (V.5), considered as a function of fiber propagation length and fiber dispersion $D$ for (a) $f_{err}^{(ISI)}=0.001$. $t_{ISI}$ for all fibers satisfy $t_{ISI} \approx 0.697$; (b) $f_{err}^{(ISI)}=0.1$ and $f_{err}^{(ISI)}=0.0001$ to display the sensitivity to the choice of $f_{err}^{(ISI)}$. For each of the pairs (same line style), the curves with lower values of $f_{max}$ correspond to $f_{err}^{(ISI)}=0.0001$ and the curves with higher values of $f_{max}$ to $f_{err}^{(ISI)}=0.1$. The photon loss due to ISI is $t_{ISI}\left(f_{err}^{(ISI)}=0.1\right) \approx 0.331$, $t_{ISI}\left(f_{err}^{(ISI)}=0.01\right) \approx 0.562$, and $t_{ISI}\left(f_{err}^{(ISI)}=0.0001\right) \approx 0.785$.

For example, consider 300 km of LEAF fiber with 0.185 dB/km fiber loss and $D = 4.25$ ps/nm/km dispersion, i.e. $\beta_2 = -5.42$ ps$^2$/km, at channel central wavelength $\lambda \sim 1550$ nm. We obtain maximum quantum bit rate $f_{max} \approx 4.31$ Gb/s. For 10 Gb/s LEAF fiber QKD link and conservative $f_{err}^{(ISI)} = 0.001$ we see, from Fig. 7a, that 50+ km distances can be achieved. Therefore, LEAF fiber may be considered as a fiber for QKD links with moderately increased quantum bit rate.

This is to be compared to 300 km Corning® Vascade® EX2000 optical fiber (used in Ref. [12]) with $D = 20.35$ ps/nm/km chromatic dispersion at 1550 nm wavelength ($\beta_2 = -25.937$ ps$^2$/km at $\lambda \sim 1550$ nm), for which $f_{max} \approx 1.97$ Gb/s, but with ~0.16 dB/km fiber loss. In order to reach $f_{max} \approx 20$ Gb/s at 300 km, while keeping other parameters the same, $|\beta_2| \sim 0.25$ ps$^2$/km ($D \sim 0.2$ ps/nm/km) is required; for $f_{max} \approx 10$ Gb/s at 300 km, $|\beta_2| \sim 1$ ps$^2$/km ($D \sim 0.8$ ps/nm/km). These estimates are summarized in Table 3.

For hybrid quantum/classical links, fiber with a small chromatic dispersion, however, brings into consideration increased nonlinear interaction of classical, high power signal with the material of the fiber, hence increased nonlinear penalties associated with the classical communication. We show in Appendix C that nonlinear penalties are under control for fiber parameters discussed within the current problem scope and fiber length below 200 km.

It is interesting to observe on the basis of Eq. (V.3), that

$$f_{max} \propto 1/\sqrt{DL} \qquad (V.6)$$

for a given central pulse wavelength $\lambda$, while maintaining current assumptions. This is a consequence of pulse elongation Eq. (III.13) and its dependence on the characteristic dispersion length $L_D$ obtained for a Gaussian temporal shape of the initial pulse, presumed above. While Eq. (V.6) gives an accurate answer for Gaussian pulse shapes, we expect that the trends indicated by it represent the reality approximately enough for other simple pulse shapes, such as Lorentzian or sech family of shapes, with no chirp, as well. As we have mentioned above, the optimal $f_{rep}$ can depend also on the duration of the input pulse. We shall turn to our discussion of this point next.

The above discussion employed an approximate evaluation of $f_{rep}$. For high quantum bit rate QKD links, say, up to 10 Gbps, the QKD performance measured by the quantum secret key rate ($R_{sec}$) and QBER, should be optimized with respect to $f_{rep}$. That is in addition to the optimization over $\mu$ in the quantum channel model presented in Section IV.B. In other words,

$$R_{sec} = R_{sec}\left(p_{ISI}\left(f_{rep}\right), t_{ISI}\left(f_{rep}\right), f_{rep}, \mu\right), \qquad (V.7)$$

needs to be optimized over $f_{rep}$ and $\mu$ for each value of the fiber propagation distance $L$ and fiber dispersion $D$ using Eqs. (III.19) and (III.20). QBER also should be considered as a function of $f_{rep}$ and $\mu$:

$$QBER = QBER\left(p_{ISI}\left(f_{rep}\right), t_{ISI}\left(f_{rep}\right), f_{rep}, \mu\right). \qquad (V.8)$$

Furthermore, we mentioned above that there is an optimal initial pulse duration for which the relative broadening caused by chromatic dispersion is the smallest. That means that the quantum secret key rate $R_{sec}$ and QBER also depend on pulse duration and should be optimized with respect to it.

In our modeling we use Eq. (V.3) to determine an approximate "ideal" quantum bit rate instead, with the limitation $f_{rep} < 10$ GHz for 10 Gb/s QKD links. For 1 Gb/s QKD links we can see from Fig. 7 that $f_{rep} = 1$ GHz is the ideal quantum bit rate. Also, we do not seek an optimal initial pulse duration value, rather use the pulse duration published in the experimental works we set to model.

**Table 3**: 300 km, up to 20 Gb/s QKD system. (In the text, we refer to the fiber families by their representative fibers in this table, even though the modeled parameters do not necessarily agree with their specification for each representative fiber.)

| Representative fiber | Dispersion @1550nm [ps/nm/km] | Dispersion slope @1550nm [ps/nm$^2$/km] | Modeled loss [dB/km] | $f_{max}$ [Gb/s] |
|---|---|---|---|---|
| Corning® LEAF® optical fiber | 4.25 | 0.085 | 0.185 | 4.31 |
| Corning® Vascade® EX2000 optical fiber | 20.35 | 0.06 | 0.16 | 1.97 |
| Low dispersion fiber #1 | 0.8 | 0.06 | 0.16 | 10 |
| Low dispersion fiber #2 | 0.2 | 0.06 | 0.16 | 20 |

In our modeling we use Eq. (V.3) to determine an approximate "ideal" quantum bit rate instead, with the limitation $f_{rep}$ < 10 GHz for 10 Gb/s QKD links. For 1 Gb/s QKD links we can see from Fig. 7 that $f_{rep}$ = 1 GHz is the ideal quantum bit rate. Also, we do not seek an optimal initial pulse duration value, rather use the pulse duration published in the experimental works we set to model.

### C. Role of quantum bit rate on the secret key rate ($R_{sec}$) and QBER

With the goal of comparing the results of our model with experiments, we carried out modeling of QKD links with varying quantum detector dead time $10^{-6}$ - $10^{-10}$ s for 1 and 10 Gb/s QKD system using APD and SNSPD, respectively. We included the cases with quantum bit rate $f_{rep}$ depending on distance as in Eq. (V.3), but limited it by maximum 1 or 10 Gb/s, respectively. We compared fiber with the set of parameters shown in Table 2 (the LDF can be considered as a hypothetical fiber).

In the examples below and in all the figures through the rest of the paper, we plot the 853 b/s secure key rate threshold to keep the security level for 100 Gb/s classical communication channel the same as in Ref. [8] (i.e. 1 key per 60 Gb of data due to practical limitations on the refresh rate; see the discussion in Section II.A), even though it could require unrealistic refresh period shorter than 1 s. An 8.53 b/s secure key rate threshold could also have been used. Similarly, we used the 9% QBER threshold for the quantum FEC as quoted in Ref. [8], with the understanding that different FEC realization can require different QBER threshold. The 9% QBER threshold value was obtained for the CASCADE error correction algorithm to be able to distill secret bits.

**Example 1: 1 Gb/s QKD system**

In this example we study the performance of a 1 Gb/s QKD only system that is meant to mimic the one experimentally implemented and described in Ref. [11]. Our goal is to calculate the same system performance characteristics as experimentally observed for the system parameters as close as those published in Ref. [11]. The parameters used in our simulations are listed in the captions to the figures or in the figures themselves.

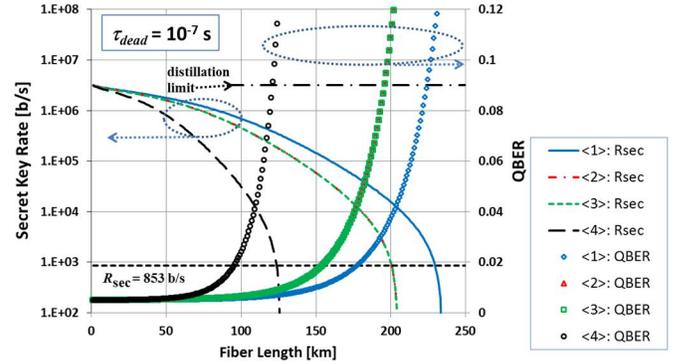

**Figure 8.** Secure key rate vs. link distance. **Fixed $f_{rep}$ = 1 Gb/s and optimized $\mu$**. According to Fig. 7, maximal quantum bit rate is not limited by dispersion at any fiber length. Comparison to experimental results for SMF28e® optical fiber with 0.3 dB/km loss and differential-phase-shift QKD in Ref. [11] – Experiment: at 45 km, $R_{sec}$ ~ 293 kb/s; this figure for $\tau_{dead} = 10^{-7}$ s: at 45 km, $R_{sec}$ ~ 994 kb/s and QBER 0.53%. *The legend is identical in the figures in Sections C and D below (Figs 9 – 18). Fiber labels are listed in Table 2.* Other used parameters: $\Delta t_{gate} = 0.5/f_{rep}$, $\tau_{FWHM,0} = 0.05/f_{rep}$, $f_{err}^{(ISI)} = 0.001$. COW protocol with V = 0.997, $p'_{dc} = 1 \times 10^{-5}$ ns$^{-1}$/APD, $\eta_{duty} = 0.71$, $\rho_{AP} = 0.008$, and $\eta = 0.19$.

Figure 8 shows that, since the maximal quantum bit rate is limited to no more than 1 Gb/s, there is a little advantage to using low dispersion fibers – all considered fibers have low enough dispersion to enable $f_{rep}$ = 1 Gb/s. See Fig 7. The changes in $R_{sec}$ and QBER when using fixed average photon number $\mu$ = 0.5 and $f_{rep}$ = 1 Gb/s are also marginal in the studied case (and, as mentioned earlier, we optimize $\mu$ and $f_{rep}$ to get best $R_{sec}$, even at the expense of worse QBER).

Next, in Fig. 9 we fix $\mu$ and $f_{rep}$ according to Ref. [11], and use $\tau_{dead}$ = 500 ns to reproduce the experimental observation for the SMF28e® optical fiber used in the experiments rather closely – the comparison is shown in Table 4.

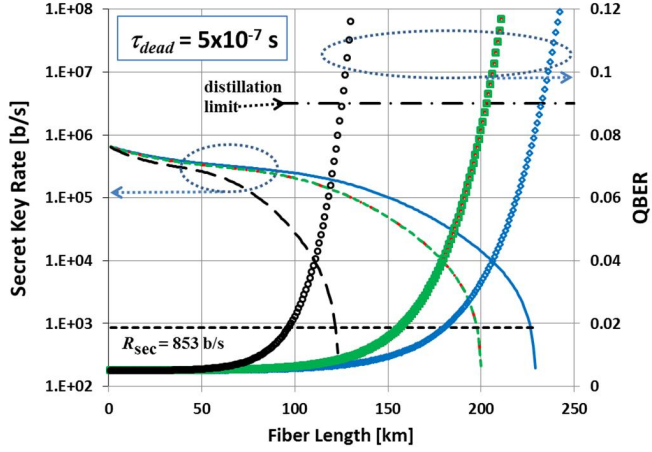

**Figure 9.** Secure key rate vs. link distance. **Fixed $f_{rep}$ and $\mu$**; $f_{rep}$ = 1 GHz, $\mu$ = 0.5. Comparison to experimental results for SMF28e® optical fiber with 0.3 dB/km loss and differential-phase-shift QKD in Ref. [11] shown in Table 4. Corning® Vascade® EX2000 optical fiber reaches ~225 km. *The legend: See Fig. 8. Fiber labels are listed in Table 2.* Other used parameters: $\Delta t_{gate}$ = 0.5/$f_{rep}$, $\tau_{FWHM,0}$ = 0.05/$f_{rep}$, $f_{err}^{(ISI)}$ =0.001. COW protocol with V = 0.997, $p'_{dc}$ = 1×10$^{-5}$ ns$^{-1}$/APD, $\eta_{duty}$ = 0.71, $\rho_{AP}$ = 0.008, and $\eta$ = 0.19.

**Table 4**: 1 Gb/s quantum bit rate performance comparison

| Fiber | Experiment, Ref. Dynes [11] | This work, $\tau_{dead}$ = 500 ns |
|---|---|---|
| QBER at 45 km | <1% | 0.54% |
| $R_{sec}$ at 45 km | 293 kb/s | 282 kb/s |

### Example 2: 10 Gb/s QKD system

In the second example we study the performance of a 10 Gb/s QKD only system that is meant to mimic the system experimentally implemented and described in Ref. [4]. Again, our goal is to calculate the system performance characteristics as observed for the experimental system.

Before investigating the correspondence of our model results with the experiment, we examine the dependence of the system performance on several parameters and fibers. In the first set of calculations, we employ optimization over average photon number $\mu$ and evaluate the ideal $f_{rep}$, with maximal quantum bit rate not larger than 10 Gb/s. The three plots in Fig. 10 show the 10 Gb/s QKD system performance for different $\tau_{dead}$ values. We observe that the performance is affected significantly by $\tau_{dead}$ choices only at the small fiber lengths, as expected. The maximal QKD link distance is limited by fiber loss and fiber dispersion. Comparison of Corning® LEAF® and LDF optical fibers having the same loss shows that smaller dispersion leads larger maximal quantum key rate. Finally, LDF/DSF, the fiber with low dispersion, performs the best with only a small margin over

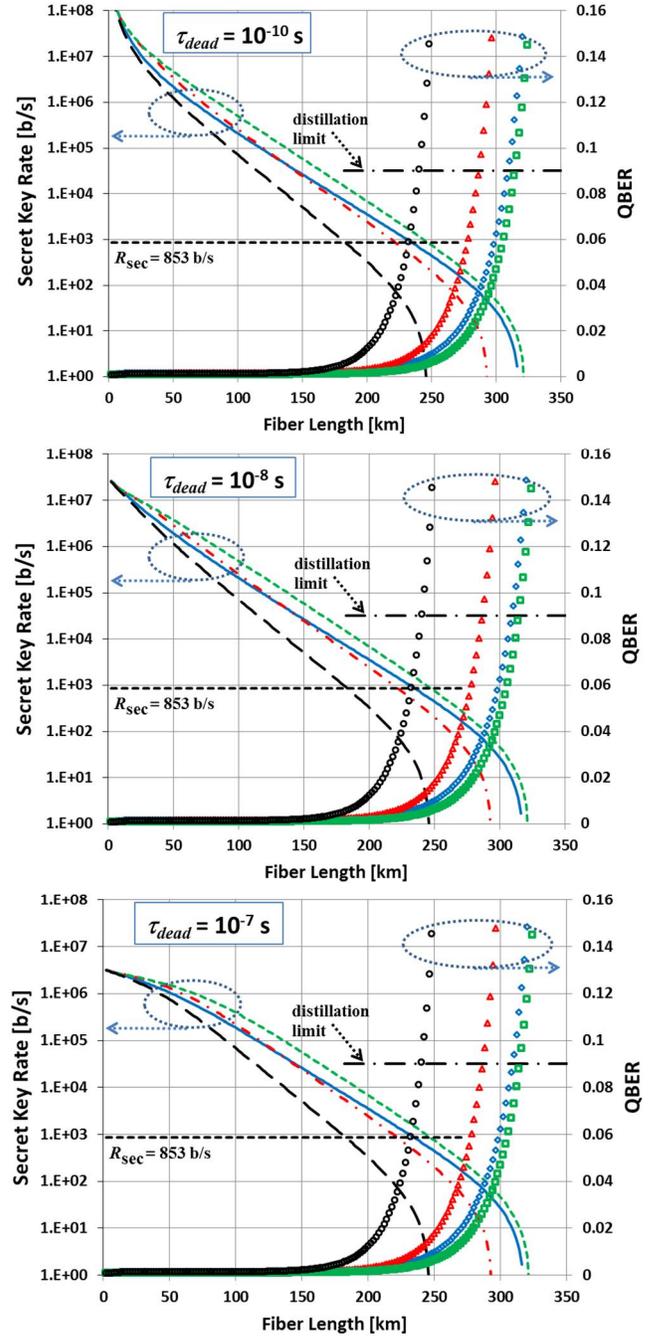

**Figure 10.** Secure key rate vs. link distance using SNSPD for small $\tau_{dead}$. **Ideal $f_{rep}$ and $\mu$ optimized**, with maximal quantum bit rate (not larger than 10 Gb/s) allowed by dispersion at each fiber length (according to Fig. 7 this limits significantly only the LDF's $R_{sec}$). Comparison to experimental results for SMF28e® optical fiber and differential-phase-shift QKD in Ref. [4] is shown in Table 5. *The legend: See Fig. 8. Fiber labels are listed in Table 2.* Other used parameters: $\Delta t_{gate}$ = 0.5/$f_{rep}$, $\tau_{FWHM,0}$ = 0.15/$f_{rep}$, $f_{err}^{(ISI)}$ =0.001. COW protocol with V = 0.997, $p'_{dc}$ = 50×10$^{-9}$ ns$^{-1}$/APD, $\eta_{duty}$ = 0.71, and $\eta$ = 0.014; no after-pulsing penalty for SNSPD assumed.

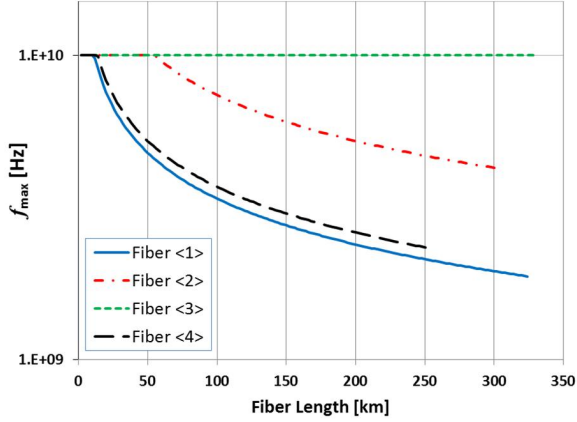

**Figure 11.** Optimized quantum bit rate *vs.* link distance for the case shown in Fig. 10. *The legend: See Fig. 8. Fiber labels are listed in Table 2.*

**Table 5**: 10 Gb/s quantum bit rate performance comparison. SMF28e® optical fiber. Ideal $f_{rep}$ and $\mu$ optimized.

|  | **Experiment, Ref.** [4] | **This work,** $\tau_{dead} = 10^{-10}$ -$10^{-8}$ s | **This work,** $\tau_{dead} = 10^{-7}$ s |
|---|---|---|---|
| QBER at 105 km | - | 0.2% | 0.2% |
| $R_{sec}$ at 105 km | 17 kb/s | 54.2 kb/s | 52.6 kb/s |
| QBER at 200 km | 4.74% | 1.2% | 1.2% |
| $R_{sec}$ at 200 km | 12 b/s | 326 b/s | 326 b/s |

Corning® Vascade® EX2000 optical fiber and Corning® LEAF® optical fiber. The curves in Fig 11 show $f_{max}$ for the different fibers whose performance was shown in Fig. 10.

In Fig. 12 we demonstrate the effect of keeping the mean number of photons in a pulse $\mu$ and the quantum bit rate $f_{rep}$ fixed or either optimized or being solved for to minimize the chromatic dispersion effect, respectively.

Next, we recalculate the performance with both $\mu$ and $f_{rep}$ fixed at $f_{rep}$ = 10 GHz, $\mu$ = 0.2. Figure 13 indicates that LDF/DSF distinguishes itself as a best performer, overall. Corning® LEAF® and Vascade® EX2000 optical fibers lag significantly behind. The dead time of the quantum detector again affects the performance of the QKD system at small propagation distances. Comparing Fig. 12 with Fig 10 and Fig 13 for $\tau_{dead} = 10^{-7}$ s, the results show that for COW protocol using the evaluation of the ideal $f_{rep}$ from Eq. (V.3) leads to more dramatic changes in performance than optimization of the mean number of photons in a pulse.

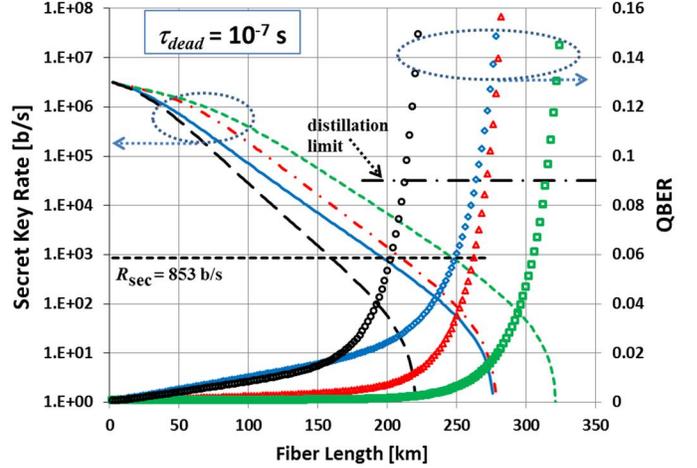

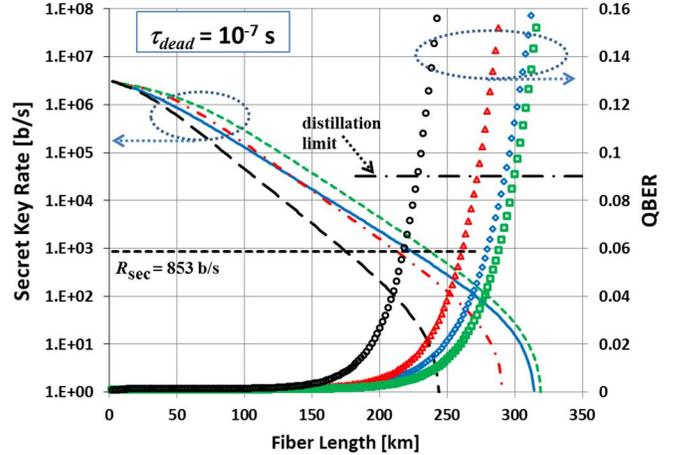

**Figure 12.** Same as Fig. 10 for $\tau_{dead} = 10^{-7}$ s, but with (top) $f_{rep}$ set to 10 GHz and with $\mu$ optimized; (bottom) with $\mu$ set to 0.2 and $f_{rep}$ chosen by solving Eq. (V.3). *The legend: See Fig. 8. Fiber labels are listed in Table 2.*

At last, we turn our attention to the comparison of our model results with the experiment. We calculate the performance with $\mu$ and $f_{rep}$ fixed according to Ref. [4]. For the choice of $\tau_{dead}$ = 300 ns, we reproduce the experimental observation rather closely – the comparison is shown in Table 6.

**Table 6**: 10 Gb/s quantum bit rate performance comparison. SMF28e® optical fiber. $f_{rep}$ = 10 GHz, $\mu$ = 0.2.

|  | **Experiment, Ref.** [4] | **This work,** $\tau_{dead}$ = 300 ns |
|---|---|---|
| QBER at 105 km | - | 0.9% |
| $R_{sec}$ at 105 km | 17 kb/s | 14 kb/s |
| QBER at 200 km | 4.74% | 8.4% |
| $R_{sec}$ at 200 km | 12 b/s | 37 b/s |

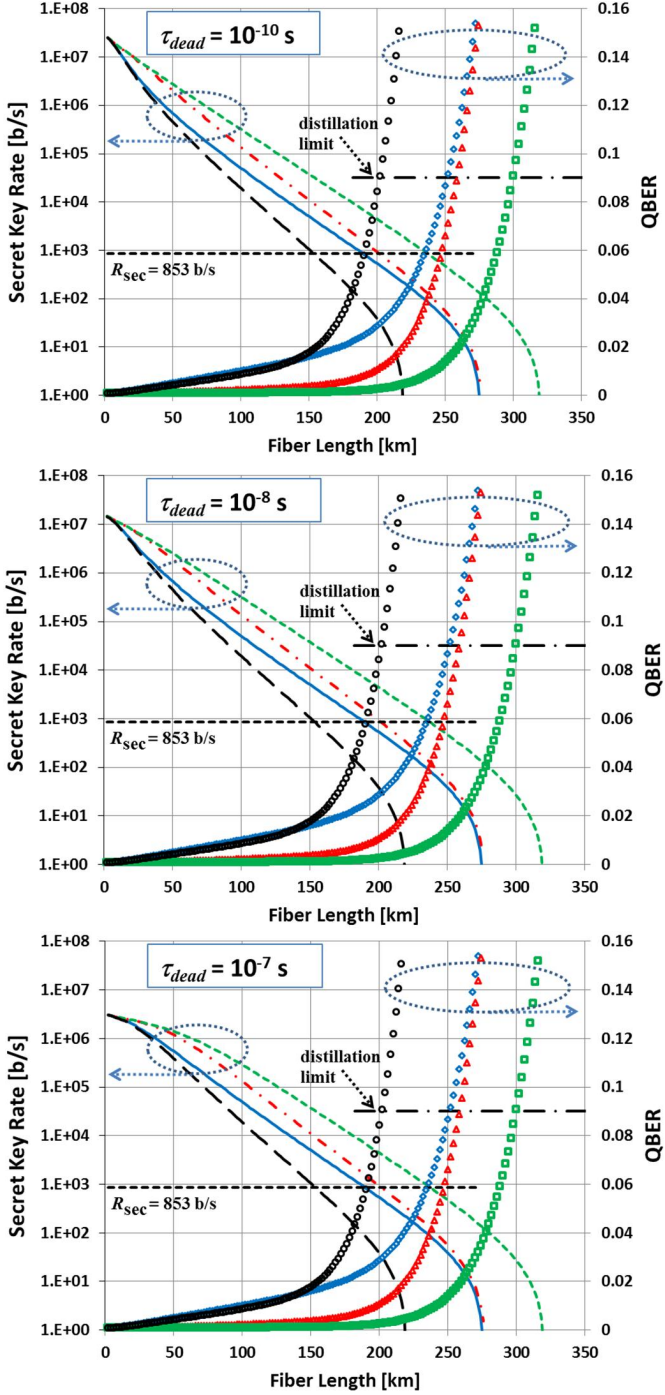

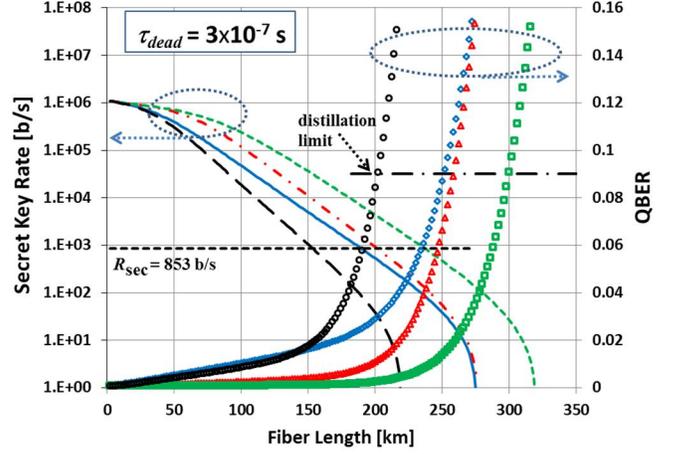

**Figure 13.** Secure key rate vs. link distance using SNSPD for small $\tau_{dead}$. **Fixed $f_{rep}$ and $\mu$**: $f_{rep}$ = 10 GHz, $\mu$ = 0.2. *The legend: See Fig. 8. Fiber labels are listed in Table 2.* Other used parameters: $\Delta t_{gate}$ = $0.5/f_{rep}$, $\tau_{FWHM,0}$ = $0.15/f_{rep}$, $f_{err}^{(ISI)}$ =0.001. COW protocol with V = 0.997, $p'_{dc}$ = $50\times10^{-9}$ ns$^{-1}$/APD, $\eta_{duty}$ = 0.71, and $\eta$ = 0.014; no after-pulsing penalty for SNSPD assumed.

**Figure 14.** Secure key rate vs. link distance using SNSPD. **Fixed $f_{rep}$ and $\mu$**: $f_{rep}$ = 10 GHz, $\mu$ = 0.2. $\tau_{dead}$ = $3\times10^{-7}$ chosen to reproduce the secure rate behavior for SMF28e® optical fiber at small distances in Fig. 5, Ref. [4]. Comparison to experimental results for SMF28e® optical fiber and differential-phase-shift QKD in Ref. [4] is shown in Table 6. *The legend: See Fig. 8. Fiber labels are listed in Table 2.* Other used parameters: $\Delta t_{gate}$ = $0.5/f_{rep}$, $\tau_{FWHM,0}$ = $0.15/f_{rep}$, $f_{err}^{(ISI)}$ =0.001. COW protocol with V = 0.997, $p'_{dc}$ = $50\times10^{-9}$ ns$^{-1}$/APD, $\eta_{duty}$ = 0.71, and $\eta$ = 0.014; no after-pulsing penalty for SNSPD assumed.

### D. Classical and quantum channels over one fiber: Dependence on the number of classical channels; 1 Gb/s QKD

One can transmit a combination of a number of classical and quantum channels over a fiber. SpRS is a constricting factor for QKD performance for a non-zero number of the classical channels being transmitted over 100+km distances along a fiber together with quantum channels. It is well known that small dispersion fiber leads to increased non-linearity impact on the system performance [29]. In Appendix C, it is shown that the nonlinear effects are negligible for LDF links with distances 200 km and less. Let's note that in this paper we don't discuss extra-long hybrid quantum/classical links with distances longer than 200 km. This issue deals with the fact that very long classical links require special methods like Raman amplification that are incompatible with quantum signals [38],[39].

Use of more classical channels, according to our current understanding, would dramatically decrease the number of well performing quantum channels in the WDM system down to 5-6 maximum (estimated based on the width of the smallest Raman scattering cross-section on the anti-Stokes side in Fig.1 and for 100 GHz compatible channel spacing), or decrease the propagation length. In order to further lower the SpRS effect, we also suggest the use of FEC for classical channels that enables a decrease in the classical signal power over the QKD link for relatively short distances. Furthermore, the improved sensitivity of modern coherent-based classical signal detectors relaxes the requirements on the channel isolation that WDM filters should provide in order not to leak any photons from the classical channel to the quantum channel. For example, this allows the WDM filters to provide isolation of ~64 dB for adjacent channels (assuming quantum detector dark count $5\times10^{-6}$/APD, quantum detector efficiency 0.07).

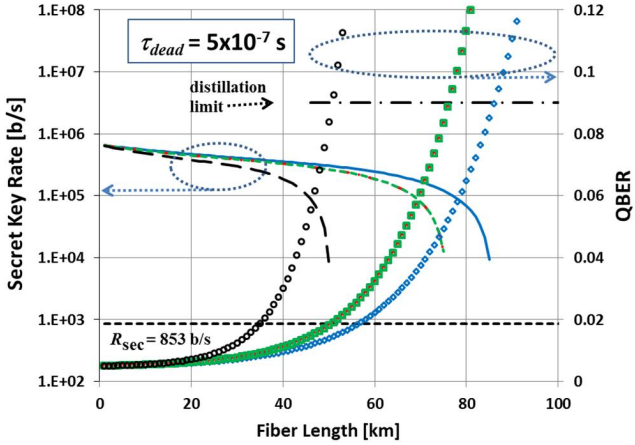
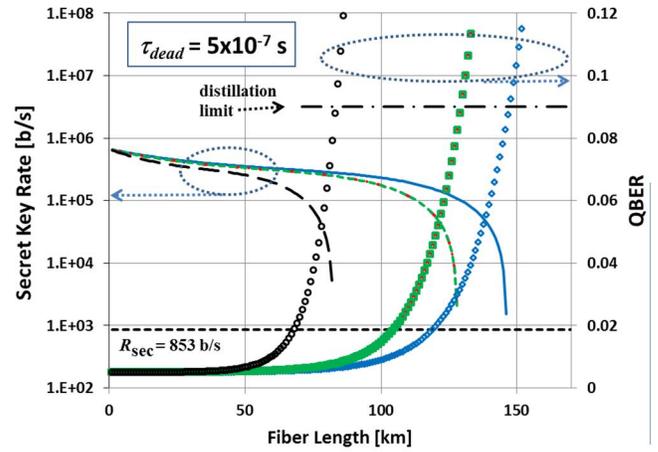

**Figure 15.** 1 Gb/s QKD performance in the presence of 2 bi-directional classical 1 Gbaud OOK channels. Corning® Vascade® EX2000 optical fiber reaches ~85 km. *The legend: See Fig. 8. Fiber labels are listed in Table 2.*

**Figure 17.** 1 Gb/s QKD performance in the presence of 4 bi-directional classical 10 Gbaud PM-BPSK channels (equivalent to 2 channel PM-QPSK, or 1 channel PM-16QAM). Compare to Fig. 9 (the effect of SpRS captured in Fig. 6 for this case). Corning® Vascade® EX2000 optical fiber reaches ~148 km. *The legend: See Fig. 8. Fiber labels are listed in Table 2.*

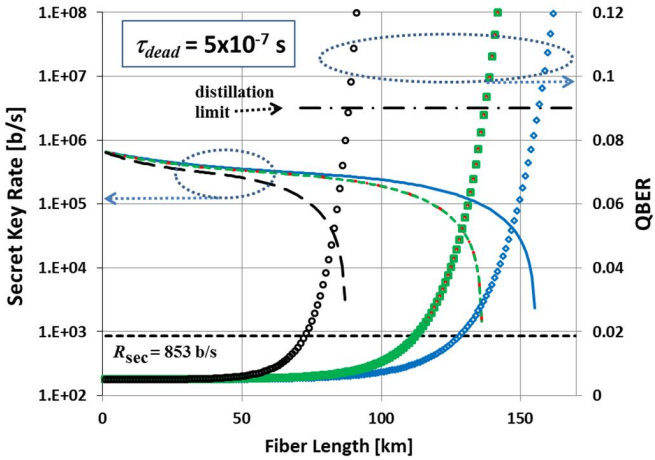
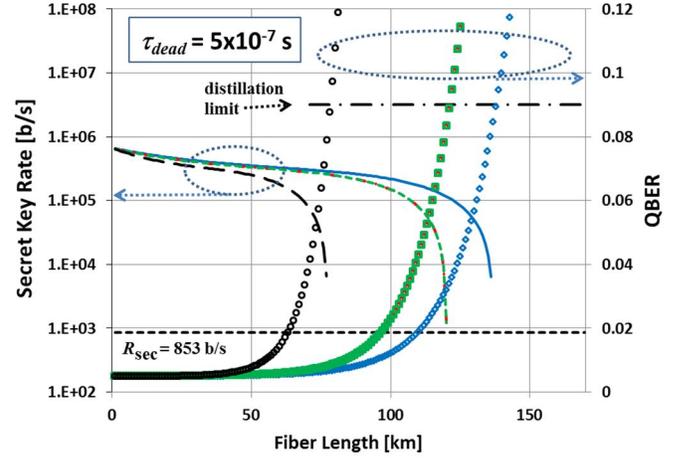

**Figure 16.** 1 Gb/s QKD performance in the presence of 2 bi-directional classical 10 Gbaud PM-BPSK channels (equivalent to 1 channel PM-QPSK). Compare to Fig. 9 (the effect of SpRS captured in Fig. 6 for this case). Corning® Vascade® EX2000 optical fiber reaches ~155 km. *The legend: See Fig. 8. Fiber labels are listed in Table 2.*

**Figure 18.** 1 Gb/s QKD performance in the presence of 8 bi-directional classical 10 Gbaud PM-BPSK channels added (equivalent to 4 channel PM-QPSK, or 2 channel PM-16QAM). Compare to Fig. 9 (the effect of SpRS captured in Fig. 6 for this case). Corning® Vascade® EX2000 optical fiber reaches ~135 km. *The legend: See Fig. 8. Fiber labels are listed in Table 2.*

In the following we use the QKD link parameters from Example 1 above (consistent with the parameters of a QKD system used in Ref. [11], i.e. SMF28e® optical fiber loss 0.3 dB/km is assumed), with fixed $f_{rep}$ = 1 GHz and $\mu$ = 0.5 for COW protocol, to evaluate the effect of additional classical channels on this QKD system. The Figs. 15-18 show the QBER and $R_{sec}$ when two bi-directional 1 Gbaud OOK and 10 Gbaud PM-BPSK, PM-QPSK, and PM-16QAM modulated classical channels are added, respectively (1 Gbaud OOK is characterized by -28 dBm receiver sensitivity [8]). We observe a steady decrease of the QKD system reach as the receiver sensitivity requirements for the modulation formats increase. For example, the QKD system reach over Corning® Vascade® EX2000 optical fiber without the presence of classical channels can be estimated as 225 km (see Fig. 9). As we add two bi-directional classical channels for PM-BPSK, PM-QPSK, PM-16QAM, and OOK, QKD system reaches approximately 155, 148, 135, 85 km, respectively (see Table 7). For a 1Gb/s QKD system, we can clearly see the benefit of low loss fiber and no effect of fiber chromatic dispersion. We also note that the receiver sensitivity requirements enable us to evaluate, for example, the performance of QKD signals in the presence of two bi-directional PM-QPSK classical channels and 4 bi-directional PM-BPSK.

**Table 7**: 1 Gb/s QKD system reach over Corning® Vascade® EX2000 optical fiber

| Modulation format in the classical channel | Number of bi-directional classical channels | QKD system reach [km] |
|---|---|---|
| No classical channel | 0 | 225 |
| PM-BPSK 10 Gbaud | 2 | 155 |
| PM-QPSK 10 Gbaud | 2 | 148 |
| PM-16QAM 10 Gbaud | 2 | 135 |
| OOK 1 Gbaud | 2 | 85 |

### E. COW and BB84 and modulation formats: Effect of OOK and BPSK

In this section, we present the study of the $R_{sec}$ and QBER performance advantages of 10 Gbaud PM-BPSK over 1 Gbaud OOK for two QKD algorithms, BB84 and COW, with fixed $f_{rep}$ = 1 GHz, and assuming SpRS with cross-section $2.6 \times 10^{-9}$(km.nm)$^{-1}$. The results are summarized in Fig. 19. From Figs. 19(a) and (c) or 19(b) and (d), we observe that QKD system performs better when two bi-directional PM-BPSK modulated classical channels are employed than when two bi-directional OOK modulated classical channels are used. Also, COW QKD protocol outperforms BB84 in both cases of OOK and PM-BPSK classical channel modulation, see Figs. 19(a) and (b) or 19(c) and (d).

### F. Classical and quantum channels over one fiber: Dependence on the number of classical channels; 10 Gb/s QKD

In the last part of this section, we use the QKD link parameters from Example 2 above (consistent with the parameters of a QKD system used in Ref. [4] and Fig 14.) with fixed $f_{rep}$ = 10 GHz and $\mu$ = 0.2 for COW protocol, to evaluate the effect of additional classical channels on this 10 Gb/s QKD system. The Figs. 20-23 show the QBER and $R_{sec}$ when two bi-directional 1 Gbaud OOK and 10 Gbaud PM-BPSK, PM-QPSK, and PM-16QAM modulated classical channels are added, respectively (1 Gbaud OOK is characterized by -28 dBm receiver sensitivity [8]). Similarly to 1 Gb/s QKD case in Section V.D, we observe a steady decrease of the QKD system reach as the receiver sensitivity requirements for the modulation formats increase. However, the fiber chromatic dispersion has a big effect on the performance of a 10 Gb/s QKD system, with lower chromatic dispersion fibers outperforming fibers with moderately lower loss.

## VI. Discussion

The results of Section V show that both the decrease of fiber loss and decrease of dispersion is important for good performance of 10 Gb/s QKD links, while there is little effect associated with chromatic dispersion for 1 Gb/s QKD links over typical installed commercial fibers. For example, LEAF fiber with larger loss than Corning® Vascade® EX2000 optical fiber can achieve better 10 Gb/s QKD performance due to smaller chromatic dispersion. Increasing the quantum bit rate up to 10 Gb/s leads to impact of fiber dispersion on theQKD system performance. Thus, the use of low dispersion fibers is needed.

The importance of coherent modulation formats in classical channels to decrease Raman noise in quantum channels is also demonstrated. We address the need to operate classical and/or quantum communication channels in WDM configuration. In order to use WDM, the amount of SpRS needs to be controlled, as it is a main source of quantum channel(s) impairment. Our main focus is the minimization of the impact of SpRS and linear crosstalk by using of CMFs and FEC. From this perspective, one could consider the following set of directions to be taken to enable the quantum/classical channel WDM: (i) Utilizing low loss fiber to diminish the effect of SpRS on performance of quantum channel(s) caused by classical channel(s); (ii) Leveraging the FEC and CMFs to diminish the effect of SpRS on performance of quantum channel(s) caused by classical channel(s); (iii) optimizing WDM filtering of quantum channels. We observe:

(1) Fibers that can achieve a loss of ~0.16 dB/km or smaller allow an increase in the maximum distance of the QKD link; the lower the loss the smaller the SpRS impact.
(2) Fiber effective area, however, does not cause a significant difference in SpRS. It is understandable from the point of view of linearity of the SpRS noise contribution. A variety of fibers with very different effective areas and core material has levels of SpRS within 10%, as can be seen in Fig. 1.
(3) Complex modulation formats and FEC, hard-decision (HD) or soft-decision (SD), of the classical channels enable an increase of maximal distance in QKD link through smaller classical signal detector sensitivity requirements. For example, to achieve the same classical channels capacity we can employ one PM-16QAM, two PM-QPSK, or four PM-BPSK bi-directional channels. Even though the SpRS from one channel is smaller than from multiple channels, the required optical signal-to-noise ratio (OSNR) penalty of the PM-16QAM outweighs the benefit of lower SpRS. PM-QPSK/BPSK formats provide the same capacity and same combined penalty (required OSNR + SpRS), only the number of channels of PM-BPSK is twice as large. That means, if fewer number of classical channels is needed, PM-QPSK modulation format is more preferable than PM-BPSK, and *vice versa*.

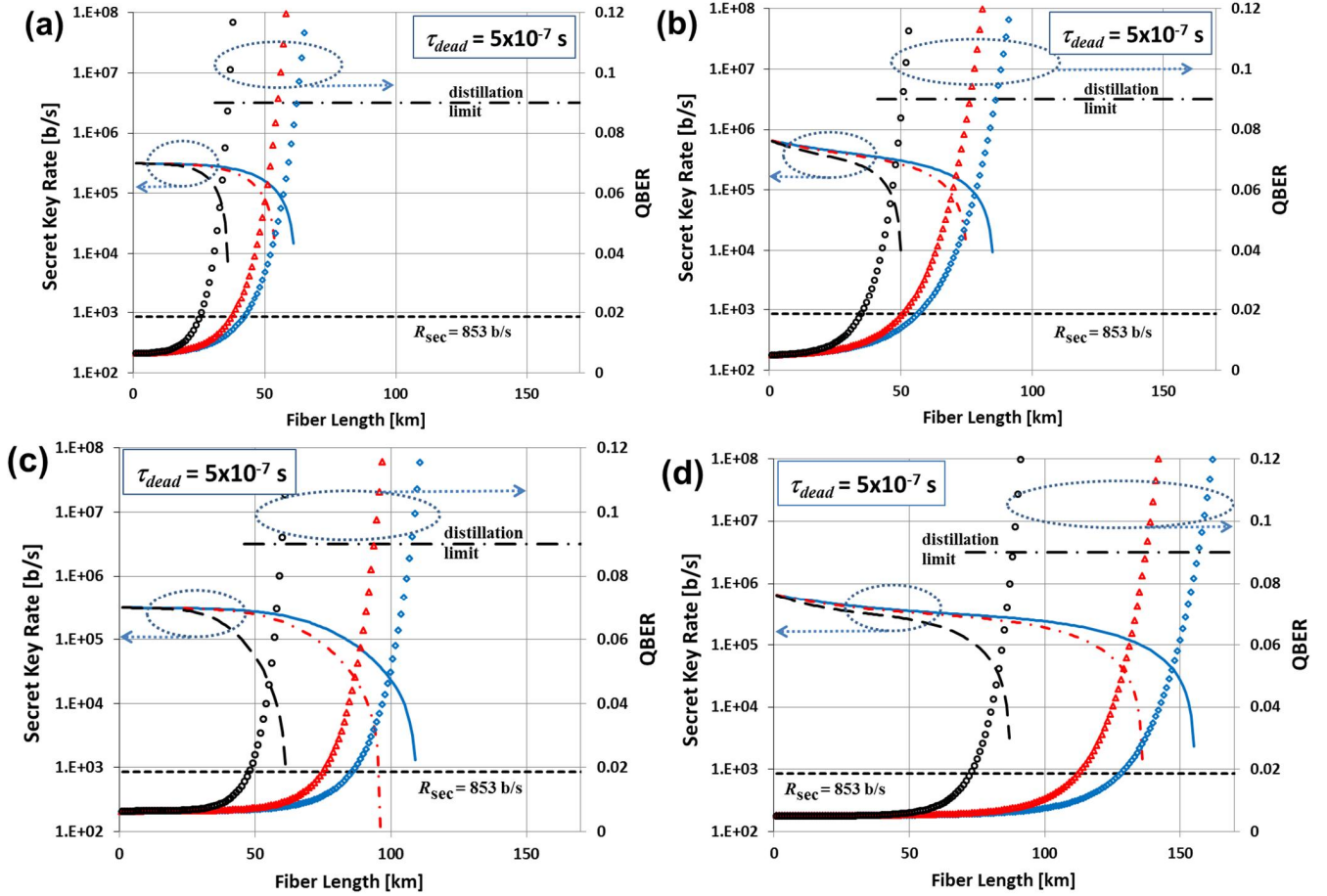

**Figure 19.** Symbols: QBER; lines: $R_{sec}$. Blue: Corning® Vascade® EX2000 optical fiber; red: Corning® LEAF® optical fiber; black: SMF28e® optical fiber. (a) BB84, 1 Gbaud OOK; (b) COW, 1 Gbaud OOK, (c) BB84, 10 Gbaud PM-BPSK; (d) COW, 10 Gbaud PM-BPSK. $\mu = 0.5$ for COW, $\mu = t_F$ for BB84.

(4) Use of more classical channels, according to our current understanding, would dramatically decrease the number of available quantum channels in the WDM or decrease the propagation length.
(5) The improved sensitivity of modern coherent-based classical signal detectors relaxes the requirements on the channel isolation provided by WDM filters.

There are ways to combat SpRS in classical/quantum WDM system that we did not include in our discussion. Namely, multicore fibers allow WDM of classical and quantum channels with smaller effect of SpRS in comparison with single core fibers, by separating the classical channels and quantum channels spatially. Under certain conditions, time division multiplexing (TDM) is also possible, with time slots determined for the use of quantum information transmission. In multicore fibers, each core can also be used for simultaneous or TDM transmission of both classical and quantum information channels.

## VII. Conclusions

We developed and extended the basic model of Eraerds *et al.* [8] to describe chromatic dispersion effects in 1 and 10 Gb/s QKD systems with and without high optical-power classical channels being present. Predictions of our simulations are supported by existing published experimental data, which gives us confidence in the accuracy of the model and its possible use to predict QKD link performance and evaluate the impact of various components, such as fiber, quantum detector, etc., or transmitter characteristics, such as quantum bit rate and pulse duration.

## VIII. Acknowledgements

The authors would like to thank Stuart Gray for providing them with Raman gain data for several optical fiber types and Bruno Huttner from ID Quantique for discussion about AES-256.

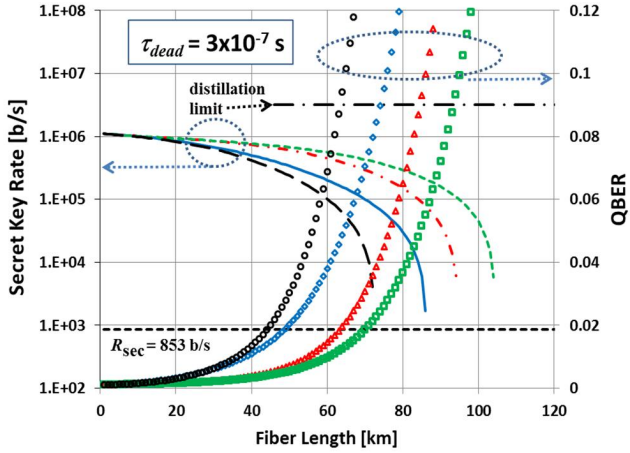

**Figure 20.** 10 Gb/s QKD performance in the presence of 2 bi-directional classical 1 Gbaud OOK channels. *The legend: See Fig. 8. Fiber labels are listed in Table 2.*

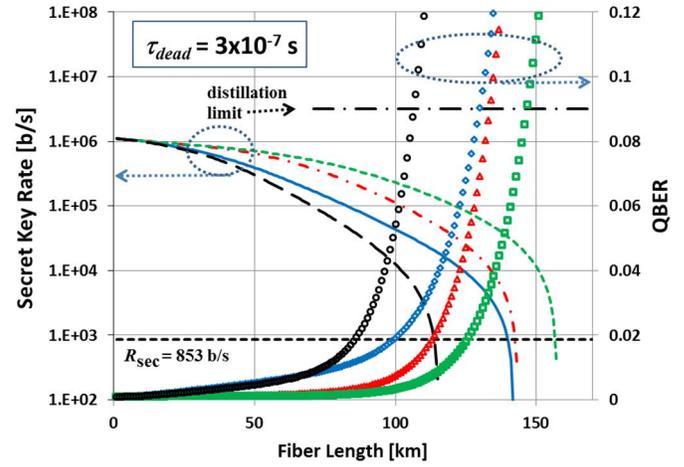

**Figure 22.** 10 Gb/s QKD performance in the presence of 4 bi-directional classical 10 Gbaud PM-BPSK channels (equivalent to 2 channel PM-QPSK, or 1 channel PM-16QAM). Compare to Fig. 14 (the effect of SpRS captured in Fig. 6 for this case). *The legend: See Fig. 8. Fiber labels are listed in Table 2.*

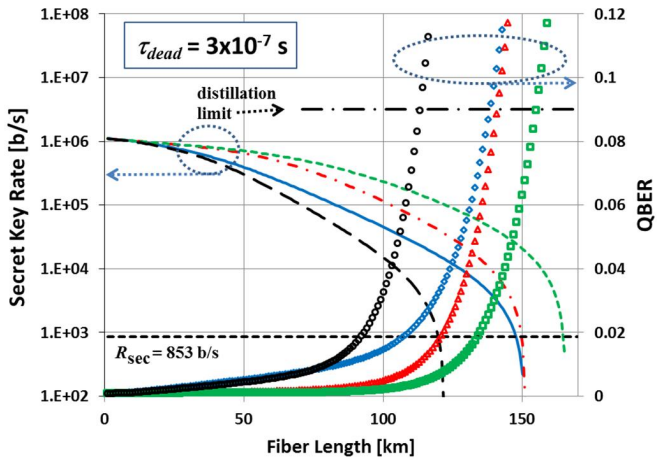

**Figure 21.** 10 Gb/s QKD performance in the presence of 2 bi-directional classical 10 Gbaud PM-BPSK channels (equivalent to 1 channel PM-QPSK). Compare to Fig. 14 (the effect of SpRS captured in Fig. 6 for this case). *The legend: See Fig. 8. Fiber labels are listed in Table 2.*

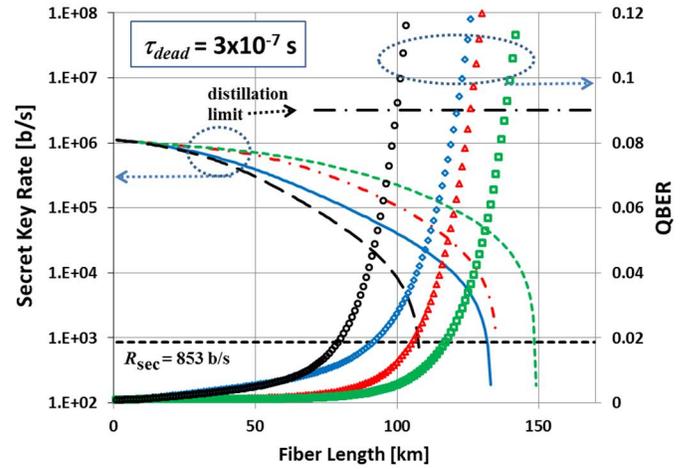

**Figure 23.** 10 Gb/s QKD performance in the presence of 8 bi-directional classical 10 Gbaud PM-BPSK channels added (equivalent to 4 channel PM-QPSK, or 2 channel PM-16QAM). Compare to Fig. 14 (the effect of SpRS captured in Fig. 6 for this case). *The legend: See Fig. 8. Fiber labels are listed in Table 2.*

# IX. Appendix

## A. BER of classical communication modulation formats

In Appendix A we briefly present the relations between BER and the received power of a classical signal in several ideal modulated formats. Assuming Gaussian noise and large signal-to-noise ratio (SNR), the theoretical dependencies BER *vs.* electrical SNR for CMFs PM-BPSK, PM-QPSK, PM-8QAM, PM-16QAM are given as [23],[40]:

$$BER_{PM-BPSK} = \frac{1}{2} erfc\left(\sqrt{SNR_m}\right), \quad (A.1)$$

$$BER_{PM-QPSK} = \frac{1}{2} erfc\left(\frac{\sqrt{SNR_m}}{\sqrt{2}}\right), \quad (A.2)$$

$$BER_{PM-SP-8QAM} = \frac{1}{2} erfc\left(\frac{\sqrt{SNR_m}}{\sqrt{5}}\right), \quad (A.3)$$

$$BER_{PM-16QAM} = \frac{3}{8} erfc\left(\frac{\sqrt{SNR_m}}{\sqrt{10}}\right), \quad (A.4)$$

where *erfc* is the complementary error function [23]. To describe the performance of a back-to-back system consisting of a transmitter and receiver, the measured dependence of BER *vs.* electrical SNR is used. In optical communication, as a rule, the optical SNR (OSNR) caused by ASE noise can be measured easily and is used to that avail: The measured electrical SNR, $SNR_m$, in this case is calculated as [40]

$$SNR_m \equiv \frac{P}{\alpha_m} = OSNR \cdot \frac{2\Delta \nu_{res}}{pB_{eq}}. \quad (A.5)$$

$P$, $\alpha_m$ are electrical signal and noise powers after detection of the optical signal, $\Delta \nu_{res}$ denotes the measurement resolution bandwidth, $p$ number of polarizations of the optical field, and $B_{eq}$ is the equivalent electrical bandwidth related to the baud rate [23]. In Ref. [34], the classical signal detector shot noise is considered as the measured noise. In this case the measured SNR is given as $SNR_m = P/\alpha_m$. Shot noise $\alpha_m$ = -58.5 dBm (noise-shot-limited condition) and Eqs. (A.1)-(A.4) were used to generate Fig. 4, left panel, which corresponds to the results in Ref. [34]. The real performance of back-to-back systems is always worse; this effect is termed "implementation penalty". The definition and mathematical model of the implementation penalty is considered next.

The technical approach to account for the difference between experimentally measured BER and ideal, noise-shot-limited, prediction of BER is to employ the notion of an implementation penalty. Implementation penalty is defined as the difference between theoretical and actual dependencies BER *vs.* OSNR in back-to-back system [41]. The example of the implementation penalty is shown in Fig. App1 for the case of PM-16QAM format. The theoretical dependence BER *vs.* OSNR for PM-16QAM modulation format is defined by Eq. (A.4), where SNR is defined by Eq. (A.5).

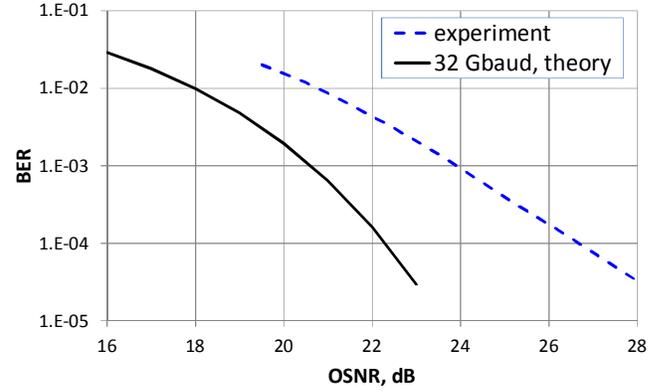

**Figure App1.** Implementation penalty of PM-16QAM back-to-back system

We saw that in the ideal system the electrical SNR is calculated by $SNR_m = P/\alpha_m$, Eq. (A.5). In the real system we may expect more complicated dependence due to nonlinearities and additional sources of noise:

$$SNR = \frac{P}{\alpha_m + \alpha_{hidden} + \beta \cdot P}. \quad (A.6)$$

The constant noise $\alpha_{hidden}$ may be caused by detection noise and noise in the local oscillator. The linear noise $\beta \cdot P$ may be caused by analog-digital converter (ADC) module and inter symbol interference due to extra filtering and, therefore, the narrowed signal spectrum. Taking into account Eq. (A.5), Eq. (A.6) may be rewritten as

$$SNR(SNR_m) = \frac{SNR_m}{1+\alpha_N + \beta \cdot SNR_m}, \quad (A.7)$$

where $\alpha_N \equiv \alpha_{hidden}/\alpha_m$ is the normalized hidden noise. Coefficients $\alpha_N$ and $\beta$ defining the hidden constant and linear noise may be found from the experimental data manually or by the least square method. The right panel in Fig. 4 was built using Eq. (A.7) with coefficient values $\alpha_N$ = 1.07 and $\beta$ = 0.0075. Actually, Fig. 4 represents out effort to rebuild Fig. 2.38 from Ref. [34] for 10 Gbaud transmission.

## B. Short pulse characteristics in linear media

Let's assume Gaussian-shaped (in time) input pulse with a linear chirp *C* given by its amplitude

$$E(z=0,t) = E_0 \exp\left[-\frac{1+iC}{2}\left(\frac{t}{\tau_0}\right)^2\right], \quad (B.1)$$

and having Gaussian-shaped input intensity

$$I(z=0,t) = |E_0|^2 \exp\left[-\left(\frac{t}{\tau_0}\right)^2\right]. \quad (B.2)$$

$\tau_0$ thus denotes the $1/e$ point of pulse's intensity. The full-width half-max (FWHM) pulse duration at the input is given by

$$\tau_{FWHM,0} \equiv \tau_{FWHM}(z=0) = 2\sqrt{\ln 2}\cdot\tau_0, \quad (B.3)$$

and after propagating a distance $L$ by

$$\tau_{FWHM,L} \equiv \tau_{FWHM}(z=L) = \tau_{FWHM,0}\sqrt{\left(1+\frac{C\beta_2 L}{\tau_0^2}\right)^2+\left(\frac{L}{L_D}\right)^2}, \quad (B.4)$$

where $L_D \equiv \tau_0^2/|\beta_2| = \tau_{FWHM,0}^2/(4\ln 2|\beta_2|)$ and $\beta_2$ is the fiber dispersion parameter. Depending on the input pulse chirp, the pulse duration can either increase or decrease initially to its minimum value

$$\tau_{FWHM,L_{min}} \equiv \tau_{FWHM}(z=L_{min}) = \frac{\tau_{FWHM,0}}{\sqrt{1+C^2}} \quad (B.5)$$

attained at $L_{min} = L_D C/(1+C^2)$, if $\beta_2 C < 0$. The final chirp $C_f$ parameter of a pulse with an initial linear chirp $C$ that propagates through a fiber of length $z$ and chromatic dispersion $\beta_2$ can be written as

$$C_f = C + \text{sgn}(\beta_2)\frac{z}{L_D}(1+C^2). \quad (B.6)$$

Thus we can obtain a $C_f = 0$ value after propagating a distance $z(C_f = 0) = -L_D \text{sgn}(\beta_2) C/(1+C^2)$, positive and the same as $L_{min}$ above when $\beta_2 C < 0$.

Fourier-transform, or bandwidth, limited pulse satisfies the time-bandwidth product (TBP) relation $2\pi\Delta\nu_{1/e}\cdot\tau_0 = 1$ or, equivalently in terms of FWHM pulse duration, $\Delta\nu_{FWHM}\cdot\tau_{FWHM,0} = 2\ln 2/\pi \approx 0.4412712$. Here, we denoted by $\Delta\nu_{1/e}$ the $e^{-1}$ intensity point of the pulse spectrum

$$I(z=0,\nu) \equiv I_0 \exp\left[-\left(\frac{\nu}{\Delta\nu_{1/e}}\right)^2\right]. \quad (B.7)$$

Using the Fourier transform of the pulse field

$$E(z=0,\nu) = \tilde{E}_0 \exp\left[-\frac{2\pi^2\nu^2\tau_0^2}{1+iC}\right], \quad (B.8)$$

in the case of no chirp, $C = 0$, we obtain

$$I(z=0,\nu) = \left|\tilde{E}_0 \exp\left[-2\pi^2\nu^2\tau_0^2\right]\right|^2 = |\tilde{E}_0|^2 \exp\left[-4\pi^2\nu^2\tau_0^2\right] \quad (B.9)$$

Comparing Eq. (B.7) and Eq. (B.9) we arrive at the TBP relation $\Delta\nu_{1/e} = 1/(2\pi\tau_0)$. FWHM pulse spectral width is then, using Eq. (B.3),

$$\Delta\nu_{FWHM} = \sqrt{\frac{1+C^2}{\pi^2}\frac{\ln 2}{\tau_0^2}} = \frac{1}{\tau_{FWHM,0}}\frac{2\ln 2}{\pi}\sqrt{1+C^2}, \quad (B.10)$$

which, for $C = 0$, results in the above TBP relation in terms of FWHM quantities.

We also mention the TBP relation in terms of the spectral widths in frequency versus the wavelength units, using $|\Delta\nu_{1/e}/\nu| = |\Delta\lambda_{1/e}/\lambda|$ and $\nu = c/\lambda$. We get $2\pi\cdot\Delta\lambda_{1/e}\cdot\tau_0\cdot c/\lambda^2 = 1$, where $\Delta\lambda_{1/e} = |\Delta\lambda_{1/e}|$. Similarly, we can write in terms of FWHM quantities $\Delta\lambda_{FWHM}\cdot\tau_{FWHM,0}\cdot c/\lambda^2 = 2\ln 2/\pi$. As a consequence, $L_D$ can be written as

$$L_D \equiv \frac{\tau_0^2}{|\beta_2|} = \frac{\tau_0}{|\beta_2|}\frac{\lambda^2}{2\pi c}\frac{1}{\Delta\lambda_{1/e}} = \frac{\tau_0}{|D|\Delta\lambda_{1/e}} = \frac{\tau_{FWHM,0}}{|D|\cdot 2\sqrt{\ln 2}\cdot\Delta\lambda_{1/e}},$$

$$\equiv \frac{\tau_{FWHM,0}^2}{4\ln 2|\beta_2|} = \frac{\tau_{FWHM,0}}{4\ln 2|\beta_2|}\frac{\lambda^2}{c}\frac{2\ln 2}{\pi}\frac{1}{\Delta\lambda_{FWHM}} = \frac{\tau_{FWHM,0}}{|D|\cdot\Delta\lambda_{FWHM}} \quad (B.11)$$

where the relation (III.14) was used.

For Fourier-transform limited pulses propagating in fiber, it is often true that $L \ll L_D$. In such case, Eq. (B.4) can be approximated as by

$$\tau_{FWHM,L}^{(C=0)} \equiv \tau_{FWHM}^{(C=0)}(z=L) = \tau_{FWHM,0}\sqrt{1+\left(\frac{L}{L_D}\right)^2}$$

$$\approx \tau_{FWHM,0}\left(1+\frac{L^2}{2L_D^2}\right) = \tau_{FWHM,0}\left[1+\frac{(|\beta_2|L)^2}{2\tau_0^4}\right]$$

$$= \tau_{FWHM,0}\left(1+\frac{8(\ln 2|\beta_2|L)^2}{\tau_{FWHM,0}^4}\right) \quad (B.12)$$

We finish this appendix by noting that pulse broadening can also be modified by higher order dispersion and making the propagating light partially coherent as a consequence of the finite linewidth of the light source [29].

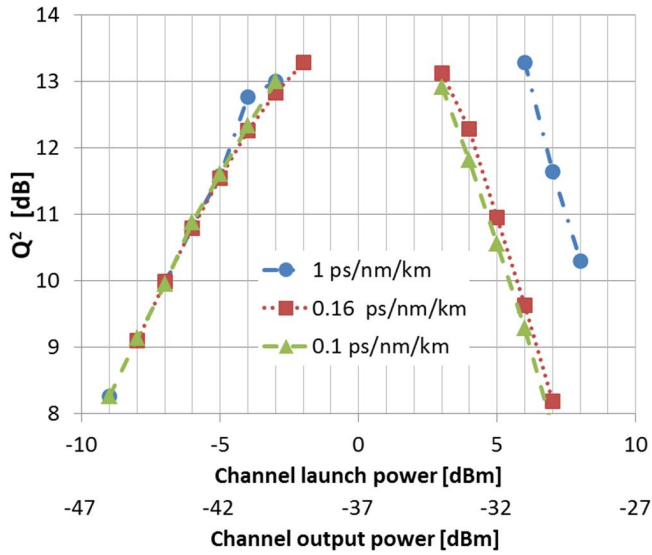

**Figure App2.** Q-factor for a nonlinear channel in a four-channel WDM PM-QPSK transmission as a function of optical power at the input and output of a fiber. 200 km propagation distance and 0.185 dB/km fiber loss assumed. Chromatic dispersion: 0.1 – 1 ps/nm/km.

### C. LDF nonlinearity

It is well known that small dispersion fiber leads to increased non-linearity impact on the system performance [29]. Hence, we need to investigate the possibility of classical channel transmission through a long low dispersion fiber (LDF) span. To estimate transmission system performance in the presence of non-linearity, we use Monte-Carlo modeling. In this paper we concluded that QKD reach for LDF links achieves 200+ km. Our modeling results show that the distance for the classical signal transmission in unrepeatered links can reach 200 km in the case of four PM-QPSK modulated channels: Taking into account 0.185 dB/km fiber loss and 0.1 ps/nm/km fiber dispersion, we may estimate PM-QPSK classical channel performance; in this case, signal loss over 200 km amounts to 37 dB. For PM-16QAM the unrepeatered link reach is shorter.

We had considered, in Section IV.A, that $P_{out}$ = -47 dBm is sufficient to detect PM-QPSK signal with BER ~ $10^{-2.4}$. Then, for 200 km long fiber with 0.185 dB/km fiber loss, -10 dBm of launch power is sufficient to reach $P_{out}$ = -47 dBm. We can see, using the numerically evaluated dependence of fiber span performance on the launch power into one of the fiber channels shown in Fig. App2, that non-linear effects are negligible for four-channel PM-QPSK transmission over 200 km fiber. The nonlinearities will be more pronounced for PM-16QAM signal. From Fig. 4 we can estimate that $P_{out}$ = -39 dBm is sufficient to detect a PM-16QAM signal with BER ~ $10^{-2.4}$. The required launch power is then -2 dBm. Fig. App2 shows that for 200 km span the non-linear effect on a single channel (in a presence of three other similar classical channels) is quite small, but it leads to about 1 dB channel power penalty, which should be taken into account.